\documentclass[article]{aa} 
\usepackage{natbib}         
\usepackage{graphicx}
\usepackage{longtable}
\usepackage[switch]{lineno} 
\bibpunct{(}{)}{;}{a}{}{,} 
\usepackage{amsmath}
\usepackage{subfig}
\usepackage{xcolor}
\newcommand{\detections}{6\,100} 
\newcommand{\accord}{600}        
\newcommand{\niveau}{13}         
\newcommand{\nstarM}{1110}       

\newcommand{\ind}[1]{_{\mathrm{#1}}}
\newcommand{\diff}{\mathrm{d}}

\def\Kepler{\emph{Kepler}}

\def\numax{\nu\ind{max}}\def\nmax{n\ind{max}}
\newcommand{\np}{n\ind{p}}
\newcommand\nm{{n\ind{m}}}
\def\Dnu{\Delta\nu}
\def\red{x_\nu}
\newcommand\nup{\nu\ind{p}}\newcommand\nug{\nu\ind{g}}
\def\doun{d_{01}}

\newcommand\nmix{\mathcal{N}}
\newcommand\dnurot{\delta\nu\ind{rot}}

\def\spec{\mathcal{P}}
\def\specp{\mathcal{S}}

\def\epsilonp{\varepsilon\ind{p}}
\def\epsilong{\varepsilon\ind{g}}

\def\ddpiover{\delta ( \deltapi )\ind{over}}
\def\ddpires{\delta  ( \deltapi )\ind{res}}
\def\ddpialias{\delta( \deltapi )\ind{alias}}
\def\ddpiordre{\delta( \deltapi )\ind{order}}

\def\Msol{M_{\odot}}

\def\ng{n\ind{g}}

\def\deltapi{\Delta\Pi_{1}}
\def\deltapil{\Delta\Pi_{l}}
\def\deltaP{\Delta P}

\newcommand\Perdeb{\tau}
\newcommand\PPerdeb{P(\Perdeb)}
\newcommand{\BV}{Brunt-V\"ais\"al\"a}
\newcommand{\NBV}{N\ind{BV}}

\begin{document}
\title{Period spacings in red giants}
\subtitle{II. Automated measurement}
\titlerunning{Gravity period spacings}
\author{%
  Vrard M., 
  Mosser B., 
  Samadi R. 
} \offprints{}

\institute{ LESIA, Observatoire de Paris, PSL Research University,
CNRS, Universit\'e Pierre et Marie Curie, Universit\'e Paris
Diderot,  92195 Meudon, France; \email{mathieu.vrard@obspm.fr} }

\abstract{The space missions CoRoT and \Kepler\  have provided
photometric data of unprecedented quality for asteroseismology. A
very rich oscillation pattern has been discovered for red giants,
including mixed modes that are used to decipher the red giants'
interiors. They carry information on the radiative core of red
giant stars and bring strong constraints on stellar evolution.}
{Since more than 15,000 red giant light curves have been observed by
\Kepler, we have developed a simple and efficient method for
automatically characterizing the mixed-mode pattern and measuring
the asymptotic period spacing.}
{With the asymptotic expansion of the mixed modes, we have
revealed the regularity of the gravity-mode pattern. The stretched
periods were used to study the evenly space periods with a Fourier
analysis and to measure the gravity period spacing, even when
rotation severely complicates the oscillation spectra.}
{We automatically measured gravity period spacing for more than
\detections\ \Kepler\ red giants. The results confirm and extend
previous measurements made by semi-automated methods. We also
unveil the mass and metallicity dependence of the relation between
the frequency spacings and the period spacings for stars on the
red giant branch.}
{The delivery of thousands of period spacings combined with all
other seismic and non-seismic information provides a new basis for
detailed ensemble asteroseismology.}


\maketitle

\voffset = 1.2cm

\section{Introduction\label{introduction}}

Using the data provided by the CoRoT satellite and four years of
observation of the space mission \Kepler, many important studies
have been carried out \citep{2009Natur.459..398D,2010ApJ...713L.176B,2011Sci...332..205B, 2011Natur.471..608B, 2012Natur.481...55B, 2012A&A...548A..10M}.
The observed pulsations correspond mostly to pressure modes which
are the signature of acoustic waves stochastically excited by
turbulent convection in the outer layers of the star. For red
giants, the radial pressure mode pattern is now understood in a
canonical form, called the universal red giant oscillation pattern
\citep{2011A&A...525L...9M}, which includes the asymptotic
contribution of the  rapid variation of the sound speed at the
second helium ionization zone \citep{2015A&A...579A..84V}. In
combination with effective temperatures, the information derived
from the radial modes is used to deliver unique information on the
stellar masses and radii \citep[e.g.,][]{2010A&A...522A...1K}. 

Red giant oscillation spectra also exhibit mixed modes. They were identified in red giants by \citet{2011Sci...332..205B}. Because they behave as acoustic waves in the envelope and as gravity waves in the core, they carry unique information on the physical conditions inside the stellar cores. Dipole mixed modes were used to distinguish core-helium burning giants (clump stars) from hydrogen-shell burning giants (RGB: Red Giant Branch stars) \citep{2011Natur.471..608B,2011A&A...532A..86M,2013ApJ...765L..41S}. Contrary to pressure modes, which are evenly spaced in frequency, and to the pattern of gravity modes, evenly spaced in period, mixed modes show a more complicated spectrum. However, their oscillation pattern can be asymptotically described \citep{1989nos..book.....U,2012A&A...540A.143M,2014MNRAS.444.3622J}. This description is based on the asymptotic period spacing $\deltapil$. The asymptotic value is defined by the integration of the \BV\ radial profile $\NBV$ inside the radiative inner regions $\mathcal{R}$. For $\ell = 1$ modes, it writes

\begin{equation}
   \deltapi = \frac{2\pi^{2}}{\sqrt{2}}\left(\int_{\mathcal{R}}
   \frac{\NBV}{r}\;  \diff r\right)^{-1}.
   \label{Brunt_equation}
\end{equation}
Its value is related to the size of the
radiative core \citep{2013EPJWC..4303002M}.

Dipole period spacing $\ell = 1$ were used to show seismic evolutionary
tracks and to distinguish the different evolutionary stages of
evolved low-mass stars, from subgiants to the ascent of the
asymptotic giant branch  \citep{2014A&A...572L...5M}. So,
identifying dipolar mixed modes is of prime importance. Furthermore, it
opens the way to measuring differential rotation in subgiants and on
the low part of the RGB
\citep{2012Natur.481...55B,2012ApJ...756...19D,2014A&A...564A..27D}
and to monitor the spinning down of the core rotation on the RGB
and in the red clump \citep{2012A&A...548A..10M}.

To date, the values of $\deltapi$ have already been extracted
manually for \nstarM\ red giant stars \citep{2014A&A...572L...5M}.
Alternatively, the method by \citet{2013ApJ...765L..41S} provides
automated estimates of the mean mixed-mode spacing but is not
intended to derive an accurate measurement of $\deltapi$. To the
contrary, the method by \citet{2015MNRAS.447.1935D}, specifically
developed for measuring asymptotic period spacings, was
presented for red giant stars where rotation is negligible, but
seems impracticable for stars showing rotational
splittings. Taking into account that \Kepler\ observed more than
15\,000 red giants and that the future ESA mission Plato may
significantly increase this number, it is then important to set up
an automatic method for measuring $\deltapi$.

In this work, we used the results obtained by \citet{2015A&A...584A..50M} to
elaborate an automated method for determining the $\deltapi$
parameter. Basically, oscillation frequencies are turned into
stretched periods that mimic the gravity periods since they are
evenly spaced. In Section \ref{Principle}, we explain the method
principle, based on this change of variable completed by a Fourier
analysis. In Section \ref{mesureDPi}, we detail the setup of the
method, including the estimate of the uncertainties. In Section
\ref{comparaison}, we compare our results with the previous
results of \cite{2014A&A...572L...5M}. This comparison helped us
to improve and speed up the new method. In Section
\ref{traitement}, we apply the method to the \Kepler\ red giant
public data; we verify the structure of the seismic evolutionary
tracks and unveil their mass and metallicity dependance on the
RGB. Section \ref{conclusion} is devoted to conclusions.

\section{Principle\label{Principle}}

Our aim is to deliver period spacings in an automated way.
Therefore, we make use of the asymptotic properties of the period
spacings presented in a companion paper \citep{2015A&A...584A..50M}.

\subsection{Period spacings}

The observed mixed-mode frequencies of giant stars do not exhibit
the same regularity as gravity modes. However, the $\deltapi$
quantity can be retrieved from the asymptotic relation which
defines the mixed-mode pattern
\citep{2012A&A...540A.143M,2013A&A...549A..75G}. We use the
implicit relation expressed in \citet{2015A&A...584A..50M}

\begin{equation}\label{deltapi_asympt}
  \tan \pi {\nu-\nup\over \Dnu(\np)}
  =
  q
  \tan \pi {1 \over \deltapi}  \left({\displaystyle{1\over\nu}
  -\displaystyle{1\over\nug}}\right),
\end{equation}
where $\nup$ and  $\nug$ are the asymptotic frequencies of pure
pressure and gravity modes, $\Dnu (\np)$ is the
frequency difference between two consecutive pure pressure radial
modes with radial orders $\np$ and $\np+1$, and $q$ is the
coupling parameter between the pressure and gravity-wave
patterns.
\newline

The asymptotic frequencies of pure dipole pressure modes are computed using the relation described by \citet{2011A&A...525L...9M}, which is called the universal pattern,

\begin{equation}\label{eqt_univ_pattern}
    \nup
     =   \left(\np + {1\over 2} + \epsilonp + \doun + {\alpha\over 2} (n-\nmax)^2 \right)\Dnu
     ,
\end{equation}
where $\epsilonp$ is the asymptotic offset, $\doun$ is the small separation corresponding to the distance (in units of $\Dnu$) of the pure pressure dipole mode compared to the midpoint between the surrounding radial modes, $\nmax = \numax/\Dnu-\epsilonp$ is the non-integer order at the frequency $\numax$ of maximum oscillation signal, and $\alpha$ is a term corresponding to the second order of the asymptotic expansion \citep{2013A&A...550A.126M}.

The asymptotic frequencies of pure dipole gravity modes are computed using the first-order asymptotic expansion \citep{1980ApJS...43..469T}

\begin{equation}\label{eq_modes_gravite}
    \frac{1}{\nug}
     =   \left(-\ng + \epsilong \right)\deltapi
     ,
\end{equation}
with $\ng$ the radial gravity order and $\epsilong$ the gravity offset. This parameter is sensitive to the stratification near the boundary between the radiative core and the convective envelope \citep{1986A&A...165..218P}.

Following Eq. (\ref{deltapi_asympt}), the period spacing $\deltaP$ between two consecutive mixed modes writes (see \citet{2015A&A...580A..96D} and \citet{2015A&A...584A..50M} for the full development)

\begin{equation}\label{eqt-zeta_deltapi}
  \deltaP = \zeta\ \deltapi
  ,
\end{equation}
where $\zeta$ is the function described in \citet{2013A&A...549A..75G} and
\citet{2015A&A...580A..96D} for expressing the relative
contribution of the inner radiative region to the mode inertia. Following \citet{2015A&A...584A..50M}, $\zeta$ is derived from the Equation~(\ref{deltapi_asympt}). It is defined by

\begin{equation}\label{eqt_zeta}
    \zeta
     =  \left[1+ \frac{1}{q} \frac{\nu^{2}\deltapi}{\Dnu (\np)} \frac{\displaystyle{\cos^{2} \pi \frac{1}{\deltapi}\left(\frac{1}{\nu} - \frac{1}{\nug} \right)}}{\displaystyle{\cos^2 \pi \frac{\nu-\nup}{\Dnu (\np)}}} \right]^{-1}
\end{equation}
with exactly the same parameters as in Eq. (\ref{deltapi_asympt}). Hence, following Eq. (\ref{eqt-zeta_deltapi}), $\zeta$ provides information on the nature of the mode : a value near 1 means that the mode is gravity dominated; on the contrary, pressure-dominated mixed modes correspond to local minima of $\zeta$. \cite{2014MNRAS.444.3622J} describe a similar property.

Equation~(\ref{eqt-zeta_deltapi}) emphasizes that the period spacings
$\deltaP$ between consecutive mixed modes are not constant. As a
result, the difference between $\deltaP$ and $\deltapi$ has to be
corrected in order to address the direct measurement of $\deltapi$.

\begin{figure}                 
  \includegraphics[width=8.9cm]{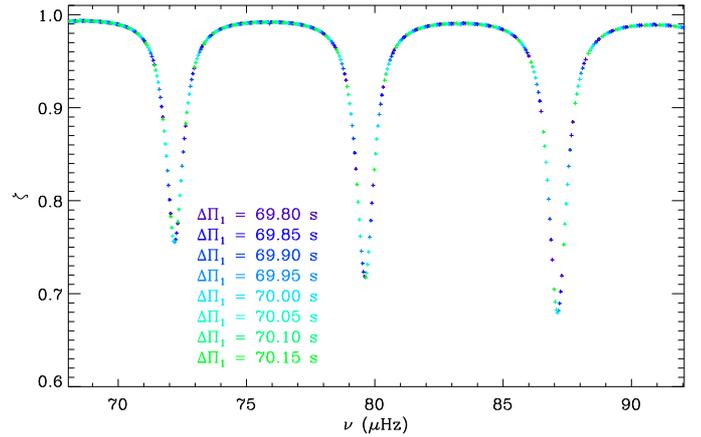}
  \caption{Precise description of the function $\zeta$ for  $\deltapi =
  70\,$s, obtained with a scan of various periods with neighboring values in
  the range $\deltapi (1 \pm \numax \deltapi / 2)$.
   \label{fig-zetacontinu}}
\end{figure}

\subsection{Stretching of the spectrum}

The main purpose of our method is to force the $\deltapi$
regularity to appear in the mixed-mode pattern. Since the
deformation of the period spacings is expressed by $\zeta$, this
function is used to modify the frequency axis of the
spectrum. In practice, each part of the frequency axis of the
spectrum is stretched according to the $\zeta$ function to account
for the difference expressed by the ratio $\deltaP/\deltapi$. We
therefore use the new variable $\tau$ defined by the differential
equation

\begin{equation}
   \diff\Perdeb = {1\over \zeta} \ {\diff\nu \over \nu^2}
   ,
   \label{stretched}
\end{equation}
where the term $\nu^{-2}$ expresses the shift from frequencies to
periods, and the term $\zeta^{-1}$ accounts for the stretching.
The role of $\zeta^{-1}$ is minor in the region of gravity-dominated mixed modes and
important in the region of pressure-dominated mixed modes. With
Eq.~(\ref{stretched}), the spectrum is reorganized as a function
of the variable $\Perdeb$, which has the dimensions of a time.

Mathematically, the change in variable corresponds to a bijection
between the frequency and period spaces, defined by

\begin{equation}
 \label{eqt-bijection}
 \mathcal{B}:
 \nu_{\nm}  
 \longmapsto  (\nm-n_0)\deltapi
 ,
\end{equation}
where $\nm$ is the mixed-mode order and $n_0$ is an arbitrary
constant. Even if the function
$\zeta$ is approximate, this bijection ensures that mixed modes will be changed into a period
comb with a distance exactly equal to the period spacing
$\deltapi$. The constant $n_0$ is arbitrary, which ensures
that the absolute numbering of the mixed modes is not necessary in order to
use Eq.~(\ref{eqt-bijection}). This avoids the difficulty of estimating
the negative gravity radial orders $\ng$ when computing
the mixed-mode frequencies with Eq.~(\ref{deltapi_asympt}) in
order to get $\nm=\ng+\np$.

We examine in the following paragraph the properties of the
function $\zeta$.

\begin{figure}[!h]                 
  \includegraphics[width=8.9cm]{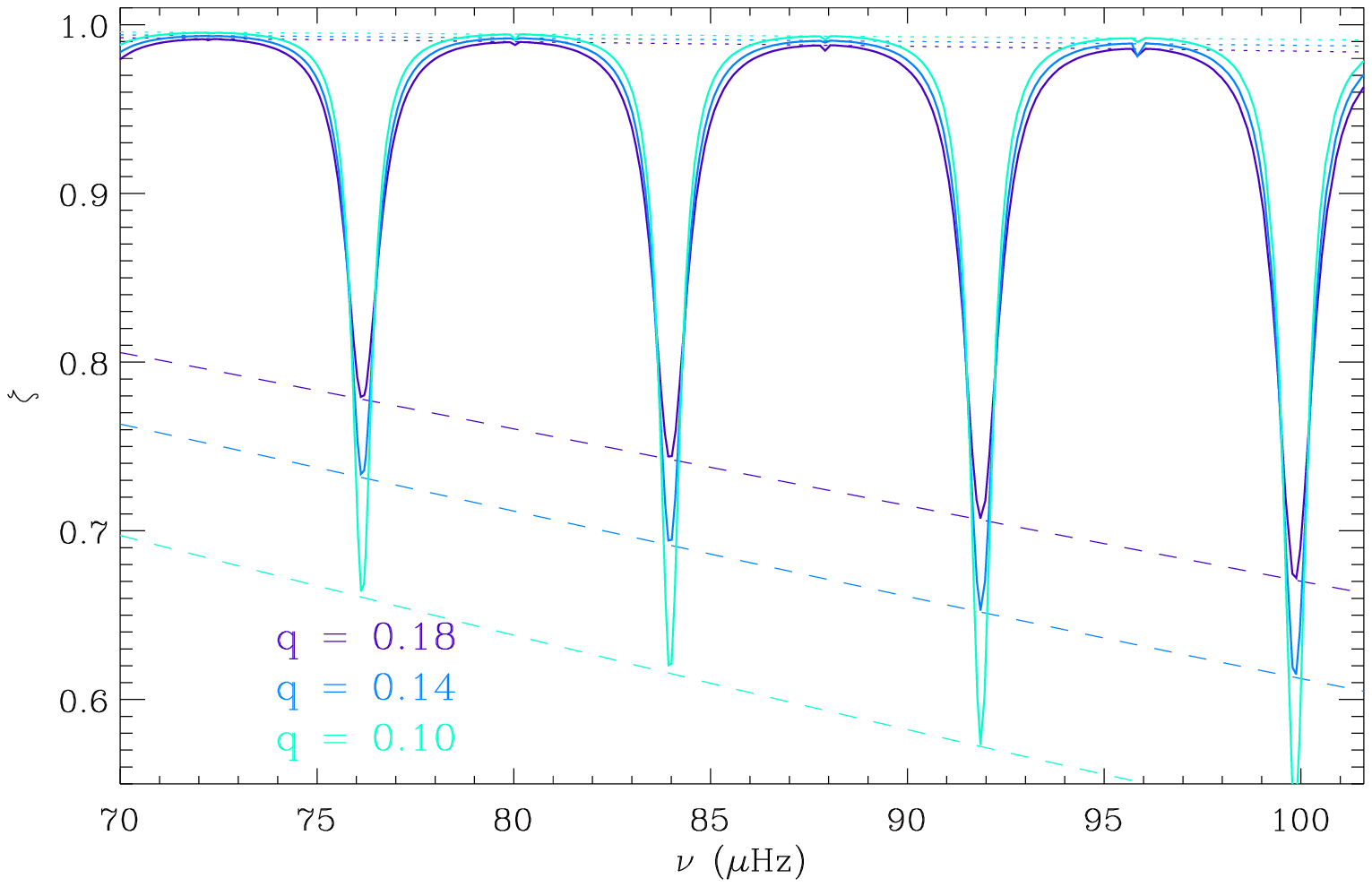}
  \includegraphics[width=8.9cm]{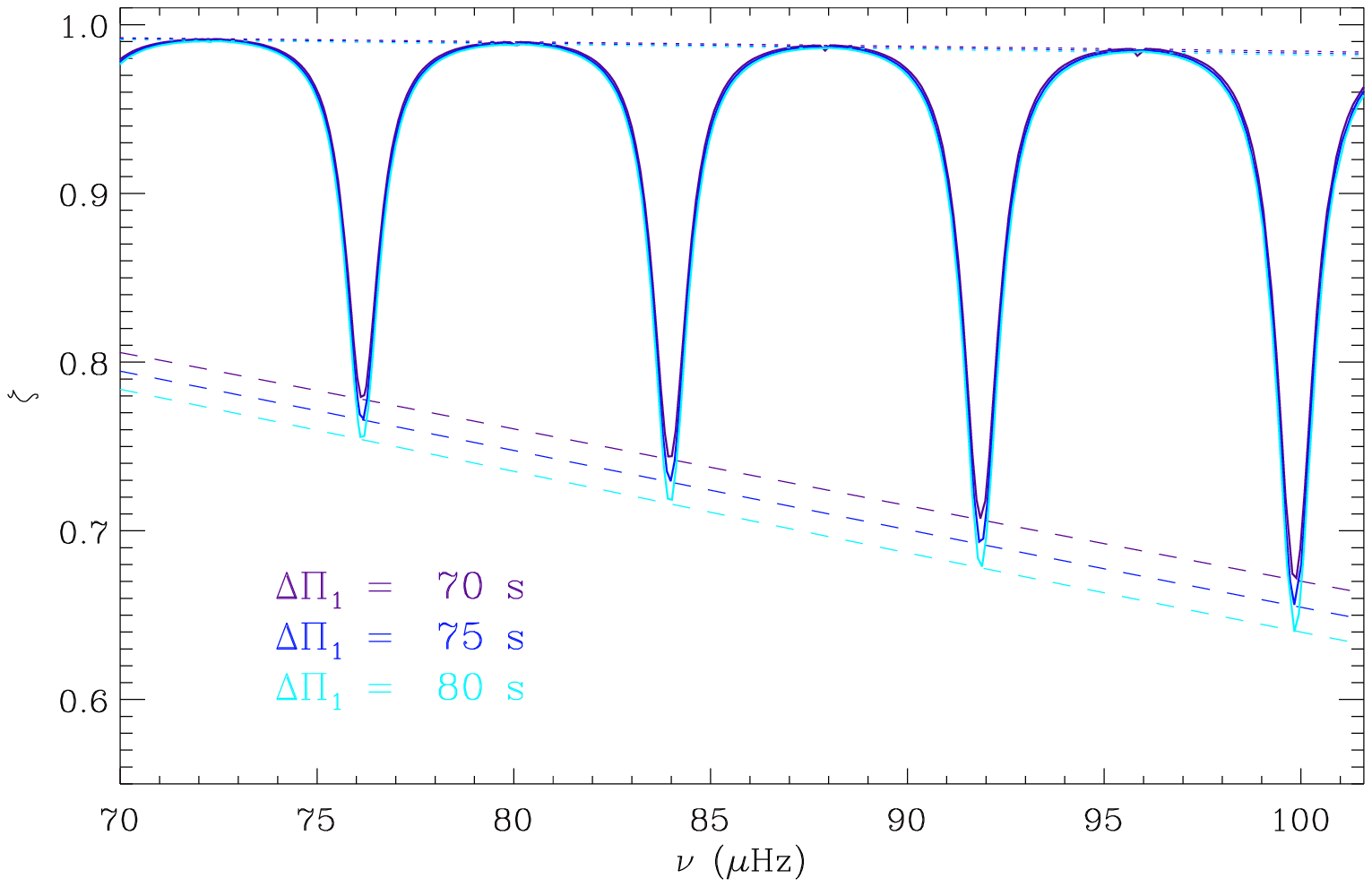}
  \includegraphics[width=8.9cm]{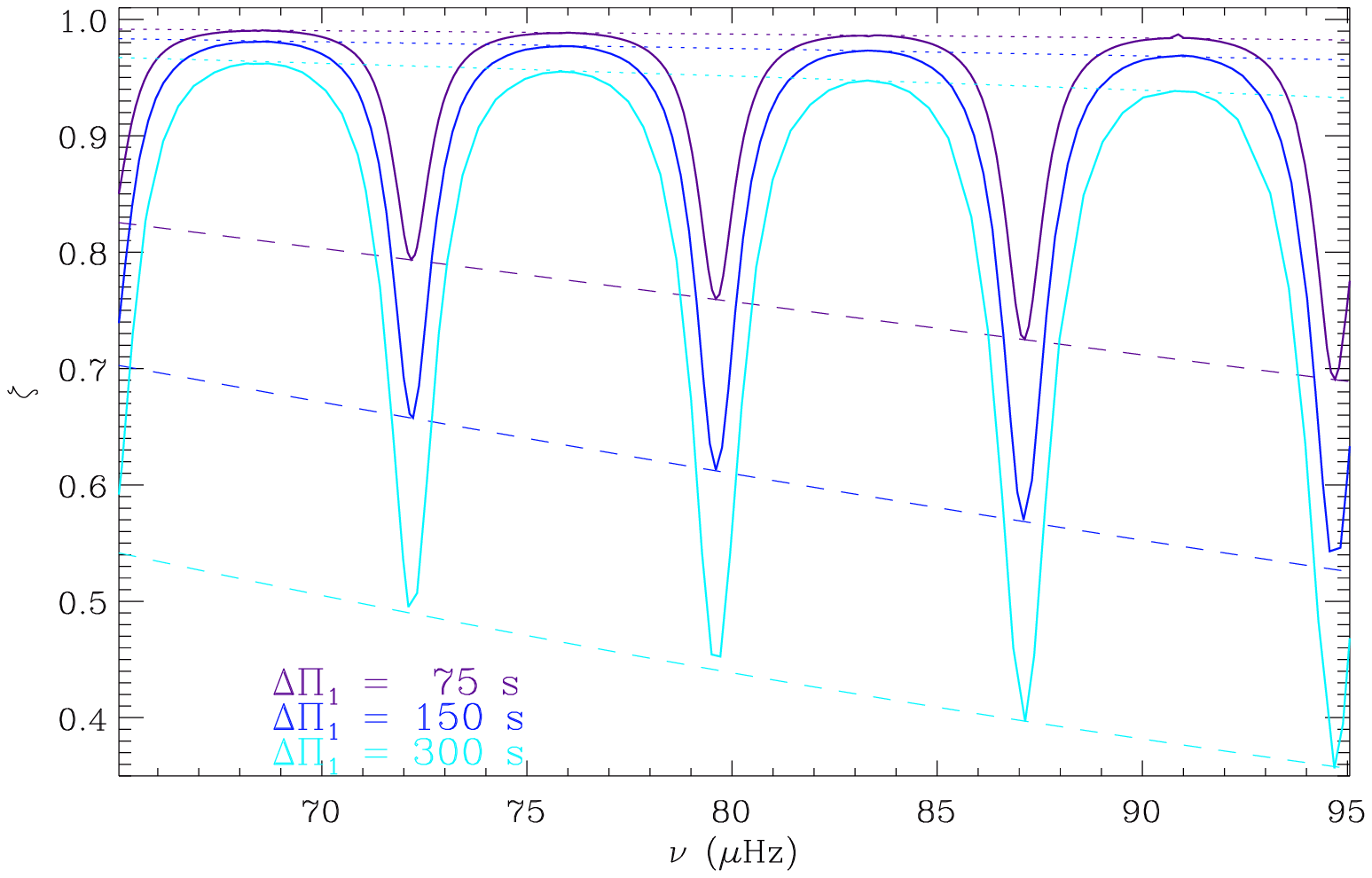}
  \caption{Function $\zeta (\nu )$ for different sets of $q$ and $\deltapi$ values
  representative of various evolutionary stages, obtained with the method
  illustrated in Fig.~\ref{fig-zetacontinu}.
  \textsl{Top:}  Typical $\deltapi$ values on the RGB, with different values of
  $q$, $\Dnu = 8\,\mu$Hz, and $\deltapi= 70\,$s. Compared to high $q$, low $q$ values correspond to deeper minima and $\zeta$
  near to 1 for gravity-dominated mixed modes.
  \textsl{Middle:}
  $\Dnu = 8\,\mu$Hz, $q=0.15$, and different values of
  $\deltapi$  found on the RGB. All curves are very similar.
  \textsl{Bottom:}  $\Dnu = 8\,\mu$Hz, $q=0.15$, and different values of
  $\deltapi$  found either on the RGB, or in the red clump. Even if large variations are seen in that
  case, the bijection (Eq.~\ref{eqt-bijection}) ensures a
  correction that is efficient enough to iterate the value of
  $\deltapi$.
  The minimum and maximum values of $\zeta$, respectively
  reached for pressure or gravity-dominated mixed modes, are
  plotted in dashed and dotted lines.
   \label{fig:fonction_debump}}
\end{figure}

\subsection{Properties of $\zeta$}

The function $\zeta$ was already implicitly depicted in previous
work \citep[e.g., Fig.~1 of ][but without the normalization by
$\deltapi$]{2011Natur.471..608B,2012A&A...540A.143M}. Here we
propose a thorough analysis of $\zeta$ and intend to show that,
even if this seems paradoxical, this function largely depends on
the seismic properties of \emph{pressure} modes and not of gravity
modes.

The functions $\zeta$ for close values of
$\deltapi$ are very similar, to those obtained for a typical RGB star (see Fig. \ref{fig-zetacontinu}). The pressure-mode parameters $\Dnu$ and  $\numax$ are fixed, whereas different values of $\zeta$ are shown for different values of $\deltapi$. This property allows us to obtain a
nearly continuous function $\zeta$, for the most precise use of
Eq.~(\ref{stretched}), with small modifications of the period
spacing around a given value of $\deltapi$. For the most efficient
computation of $\zeta$ for a given value of $\deltapi$, variations
in the range $\deltapi (1 \pm \numax \deltapi / 2)$ have to be
investigated (Fig.~\ref{fig-zetacontinu}).

The minimum values of $\zeta$ are governed by the global seismic
parameters describing the pressure mode pattern. At first order,
these minimum values are located near the first-order frequencies
of dipole pressure modes $ (\np + \epsilonp + d_{01}) \Dnu$, where
$\epsilonp$ is the asymptotic offset for pressure modes  and
$d_{01}$ is the small separation for dipole modes. The large separation ($\Dnu$)
determines the frequency difference between each minima of
$\zeta$, and $\epsilonp$ and $d_{01}$ define the position of the
dipole pressure modes \citep[e.g.,][]{2011A&A...525L...9M}, and so
determine the location of these minima. A change in these
parameters can potentially produce an important change in $\zeta$.
However, these parameters are precisely determined from the radial
mode pattern and by the frequency shift $d_{01}$ depicted by the
universal red giant oscillation pattern.

On the contrary, the function $\zeta$ hardly depends on $q$ and
$\deltapi$. As explained in \citet{2015A&A...584A..50M}, the coupling parameter
$q$ determines the depth of the minima (Fig.
\ref{fig:fonction_debump}\emph{top}). Since this parameter does
not vary much during the red giant evolution
\citep{2012A&A...540A.143M}, it has little influence on $\zeta$.
For the dependence of $\zeta$ with $\deltapi$, the scaling of
$\deltaP$ to $\deltapi$ makes the function approximately the same
for each $\deltapi$. It follows that the values of $\deltapi$ and
$q$ have only a limited impact on $\zeta$ (Fig.
\ref{fig:fonction_debump}\emph{middle}).

As a result, the determination of $\Dnu$ and of the frequency
position of the dipole pressure modes is enough to provide a
relevant estimate of $\zeta$. Therefore, using this function
to stretch the oscillation spectra results in the emergence of
a regularity corresponding to the $\deltapi$ value. Even a large
change in $\deltapi$ does not modify $\zeta$ drastically (Fig.
\ref{fig:fonction_debump}\emph{bottom}), so that a RGB star can be
treated with a clump $\deltapi$ value to initiate the
stretching, and conversely. This property is demonstrated in
Appendix \ref{ap:erreur}.

\section{Automated measurement of $\deltapi$\label{mesureDPi}}

In this section, we detail the way to measure period spacings in a
fully automated way.

\subsection{Preparation of the oscillation spectrum}

\begin{figure*}                 
  \includegraphics[width=18cm]{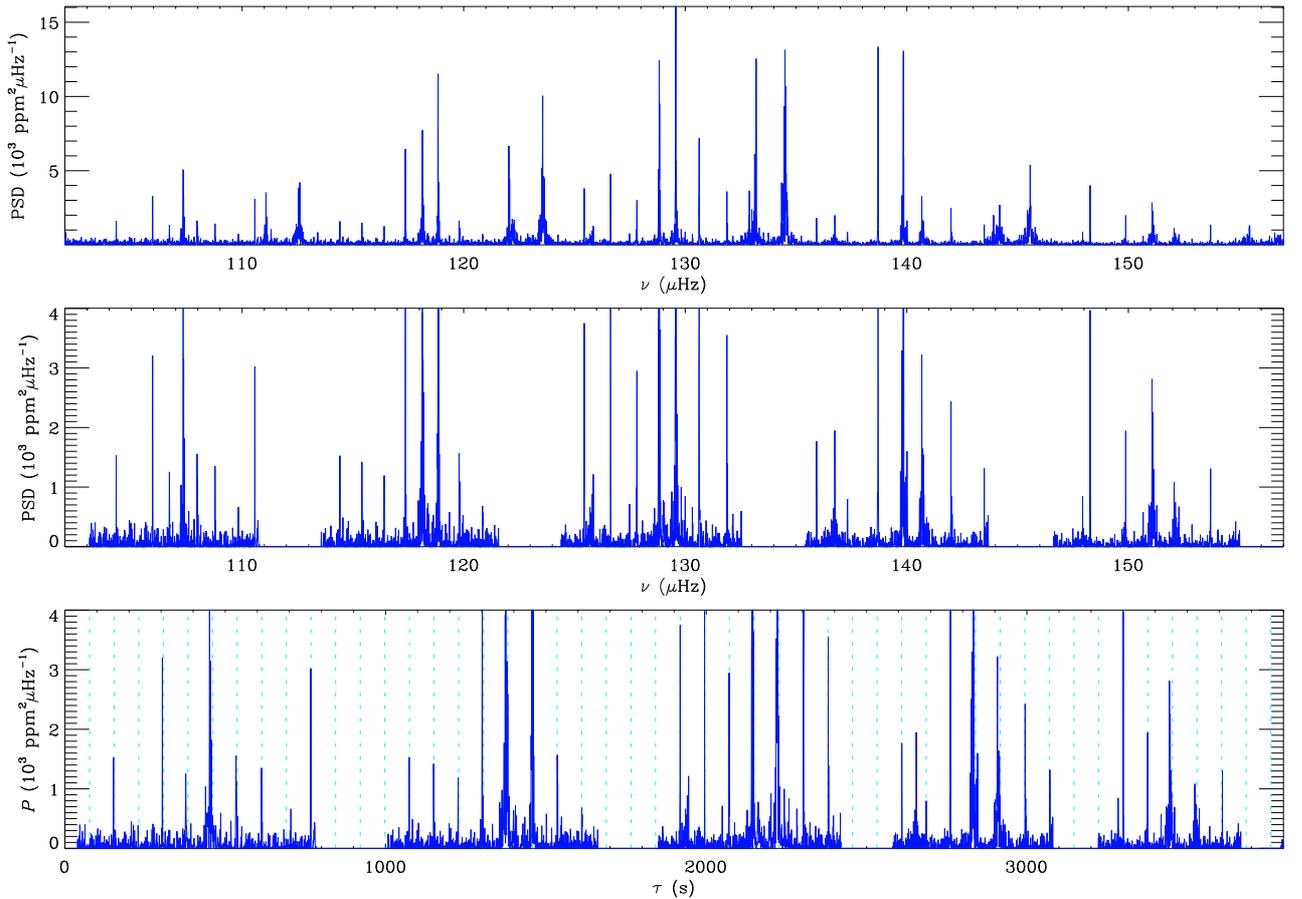}
  \caption{\textit{Top:} Oscillation spectrum of the red giant KIC9145955 in function of frequency.
\textit{Middle:} Radial and quadrupole modes have been removed
and the background has been subtracted from the spectrum.
 \textit{Bottom:} Oscillation spectrum as a function of the stretched period. The light blue
dashed lines correspond to a regular Dirac comb (indicated with
dashed lines). }
  \label{fig:spectre_debump}
\end{figure*}

The first steps of the setup are based on the pressure modes.
First estimates of the values of $\Dnu$ and $\numax$ are obtained
with the envelope autocorrelation function
\citep{2009A&A...508..877M}. These values are refined by using the
universal pattern \citep{2011A&A...525L...9M} in order to enhance
the accuracy of the determination of $\Dnu$ and to precisely
locate the different oscillation modes. We do not include
radial modes in our study since they do not exhibit mixed modes,
or quadrupole mixed modes since they are confined near the
pressure modes and do not exhibit the same pattern as dipole mixed
modes. Therefore, we suppress these modes from the spectra (second
panel of Fig. \ref{fig:spectre_debump}). In practice, this
operation only depends on the value of the large separation: we
keep part of the spectrum with a second-order reduced frequency
$\red$ verifying

\begin{equation}\label{eqt-conditions}
    \red
     = {\nu \over \Dnu}
     -\left(\np+\epsilonp +{\alpha\over 2} (n-\nmax)^2 \right)  \in [0.06,0.80]
     ,
\end{equation}
where the quadratic term accounts for the second-order asymptotic
expansion \citep{2011A&A...525L...9M,2013A&A...550A.126M}. In
order to avoid discontinuities due to the granulation background,
this operation is applied to a background-corrected spectrum
(Fig.~\ref{fig:spectre_debump}b). The background is determined as
in \citet{2012A&A...537A..30M}.

\subsection{Initial values\label{Initial}}

In order to help the  iterative process, we chose initial values of
$\deltapi$ that agree with the $\deltapi$-$\Dnu$ pattern presented
by \citet{2014A&A...572L...5M}.  For $\Dnu$ above 9.5\,$\mu$Hz, there is no ambiguity since
the only possible evolutionary status is RGB. Below this value, stars may be
on the RGB, in the clump, or leaving the clump. This last
case is equivalent to the clump phase, since the $\deltapi$ shows
continuous variation at the end of the clump. So, different cases
were then tested for $\Dnu$ below 9.5\,$\mu$Hz, with a guess value
agreeing either with the RGB or with the other evolutionary
stages.

We then stretched the oscillation spectrum using the $\zeta$
function as described in section \ref{Principle} (bottom panel of
Fig. \ref{fig:spectre_debump}).

\subsection{Spectrum of the stretched spectrum\label{Methodology}}

\begin{figure}
  \includegraphics[width=9cm]{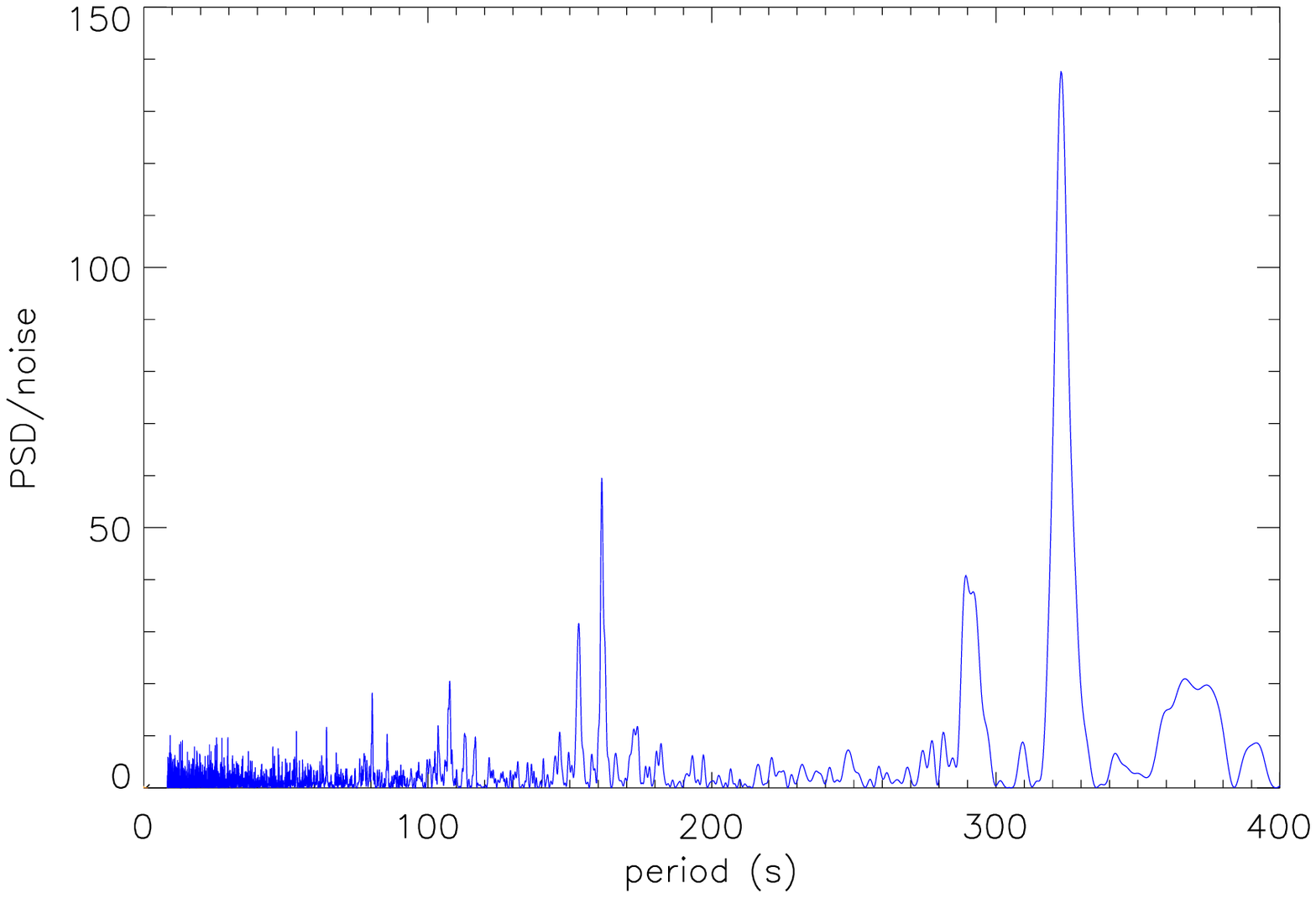}
  \includegraphics[width=9cm]{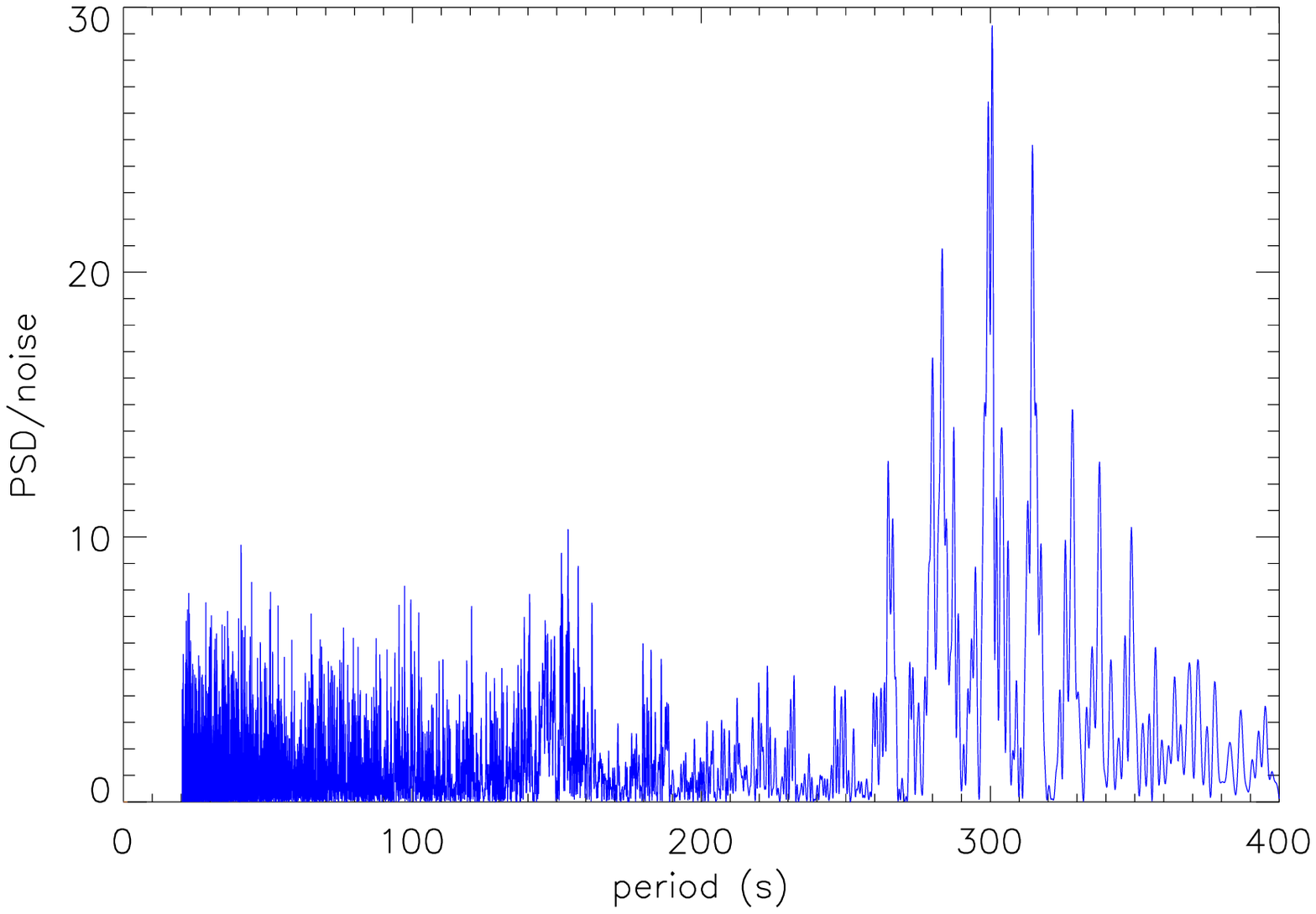}
  \includegraphics[width=9cm]{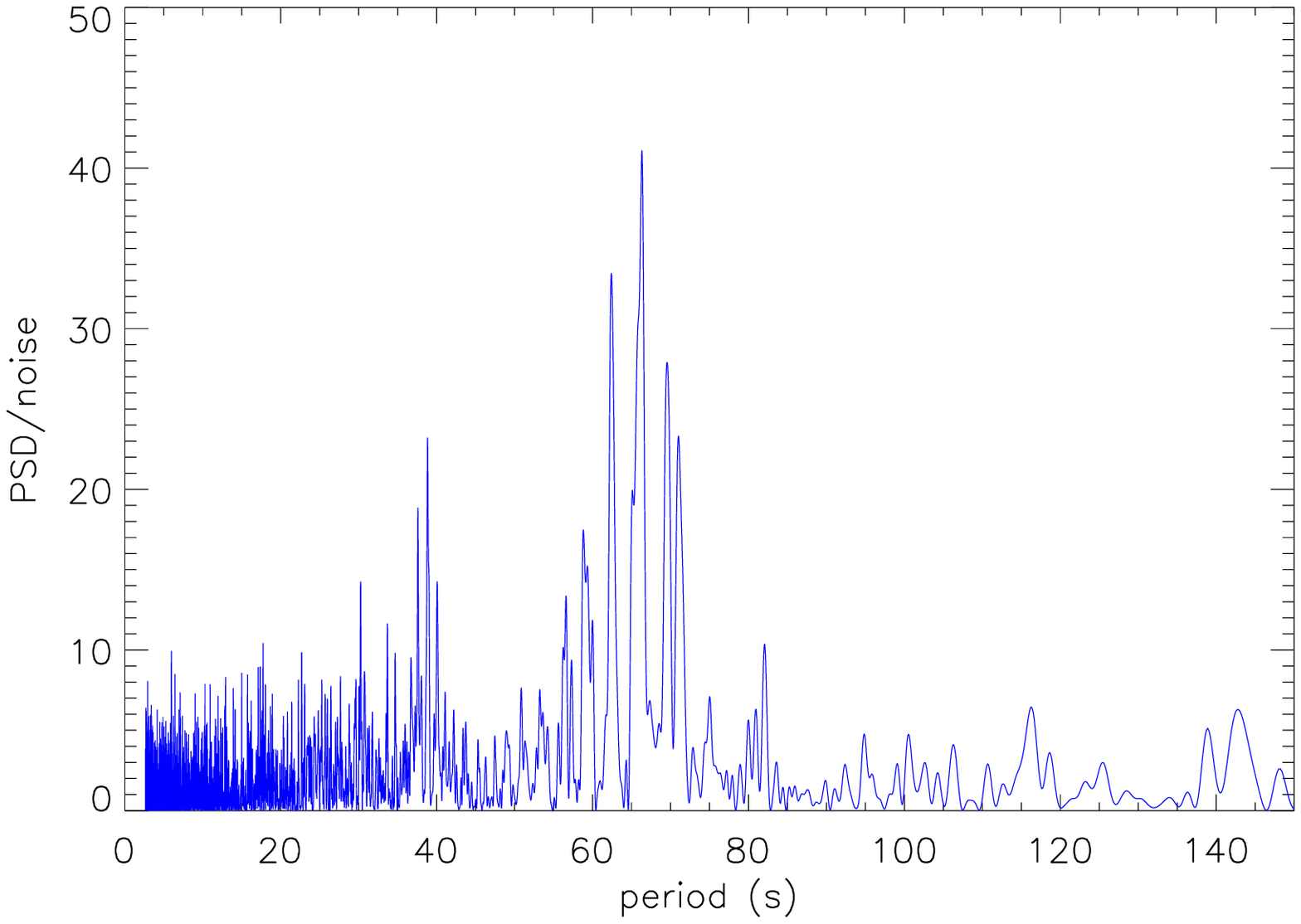}
  \caption{%
  \emph{Top}: Power spectrum of $\PPerdeb$ as a function of the period for the star
  KIC 1995859. $\deltapi$ for this star is 321\,s.
  \emph{Middle}: Same figure for KIC 1868101 with  $\deltapi=300.6$\,s.
  \emph{Bottom}: Same figure for KIC 12507577 with $\deltapi=66.3$\,s.}
  \label{fig:TF}
\end{figure}

To retrieve the $\deltapi$ value, we performed a Fourier transform
of the new spectrum $\spec (\Perdeb)$ (Fig. \ref{fig:TF}). Owing to
the form of the mixed-mode signal, with high amplitudes for the
pressure-dominated mixed modes and low amplitudes for the
gravity-dominated modes, there is no need to use a tapering
function to smooth the spectrum and reduce aliases since the
distribution of the amplitudes naturally mimics a tapering
function.

Regularity in the stretched spectrum results in a clear signature
in its Fourier spectrum (Fig. \ref{fig:TF}). As stated above, the
period signature observed is largely independent of the initial
guess value of $\deltapi$ and $q$. However, a change in these
initial values will produce a small variation in the measured
period signature. An iterative process provides a stable
measurement of the two parameters $q$ and $\deltapi$ after only four
steps (see Appendix \ref{ap:erreur}).

When necessary, we tested the different possible evolutionary
stages and kept the coherent one: when the final value of
$\deltapi$ agrees with the hypothesis on the initial value. We
also measured the mixed-mode period spacing $\deltaP$ and found
very good agreement.

\subsection{Test with a synthetic spectrum}\label{Test_synthetic}

\begin{figure}
  \includegraphics[width=9cm]{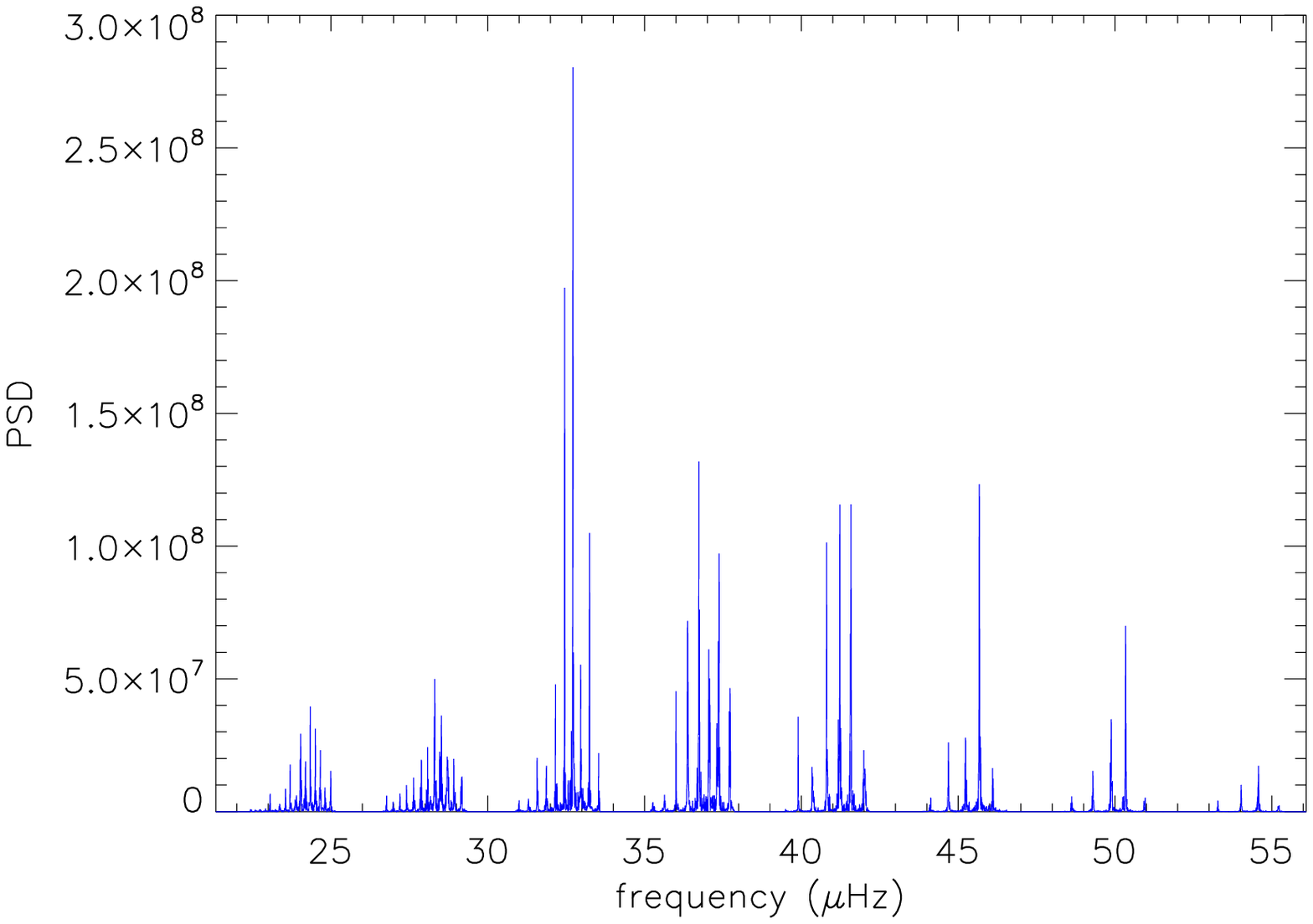}
  \includegraphics[width=9cm]{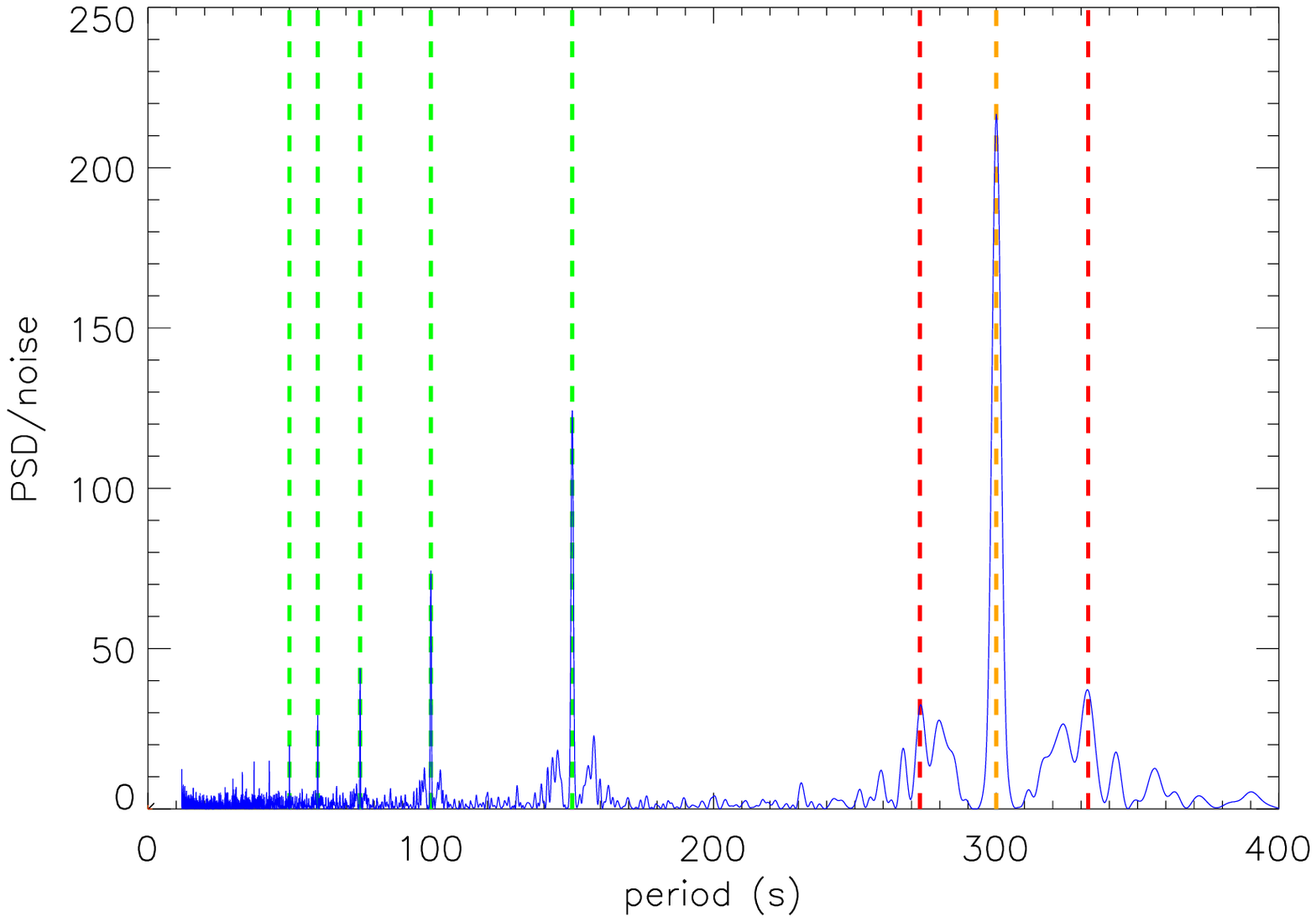}
  \caption{\textit{Top:} Simulated mixed-mode spectrum. Radial and $\ell=2$ modes are absent, as depicted in Figure \ref{fig:spectre_debump}.
\textit{Bottom:} Power spectrum of $\PPerdeb$ derived from the
asymptotic relation. The initial $\deltapi$ value was settled at
300\,s. The principal aliases observed  around $\deltapi$ are
indicated by red dashed lines. Harmonics of $\deltapi$ are also
seen (green dashed lines).
  \label{simu}
  }
\end{figure}

In order to check possible bias of the method, we performed tests
with synthetic low-degree oscillation spectra, using the
asymptotic relations for the frequencies of radial and dipole
mixed modes. Mode amplitudes were computed following
\citet{2012A&A...537A..30M}. The linewidths of the profile of the
mixed modes, described as Lorentzians, were derived from
observations. The linewidths of the radial and $\ell=2$ modes,
useless for computing the period spacing but necessary for testing
the whole automated chain, were computed following
\citet{2012A&A...540L...7B}. We finally multiplied the mixed-mode
amplitudes with a Gaussian function to take into account the amplitude
difference between pressure-dominated and gravity-dominated modes.
The FWHM of this Gaussian was fixed to one-fifth of the large
separation to match observed spectra. The result is shown in the
top part of Fig.~\ref{simu}.

We tested different values of $\deltapi$, $\numax$ and $\Dnu$ and
retrieved in each case the initial value of $\deltapi$ with a
precision much higher than 0.1\,\% (Fig.~\ref{simu}).

\subsection{Performance and uncertainties}\label{Uncertainties}

\subsubsection{Confidence level}\label{Reliability level}

To define the confidence level, we measured the mean
high-frequency noise present in the spectrum and used this value
to normalize the Fourier spectrum $\specp$ of the stretched
spectrum $\spec (\tau)$. In order to estimate the relevance of the
detection, we assumed that the statistics of $\specp$ follows a
$\chi^2$ distribution. This is not strictly the case, owing to the
rescaling from the frequency to the period domain and to the
stretching of the spectrum induced by the change of variable
(Eq.~\ref{stretched}). However, these deformations are limited in
the frequency range around $\numax$ so that we use a similar test
to the one provided by the H$_0$ hypothesis, but with a dedicated
calibration.

Assuming a $\chi^2$ distribution, the detection can be considered
reliable when the local maximum of $\specp$ is larger than ten
times the mean noise level; the detected value then corresponds to
a period signature rejecting the $H_0$ hypothesis corresponding to
pure noise with more than 99.9\,\% confidence
\citep{2009A&A...508..877M}. Simulations, consisting in retrieving
the oscillation signal in a high signal-to-noise spectrum
corrupted with white noise, showed that the detection is relevant
with the threshold level previously mentioned fixed at the value
$\niveau$.

\begin{table}[t]

\caption{Estimates of the different uncertainties for a typical RGB star with $\Dnu = 8\,\mu$Hz and  $\numax = 75\,\mu$Hz or for a typical clump star with $\Dnu = 4\,\mu$Hz and  $\numax =35\,\mu$Hz \label{table:incertitude}}

\begin{center}
\begin{tabular}{rcccc}
 \hline
                       &\multicolumn{2}{c}{RGB} &\multicolumn{2}{c}{clump} \\
        $\deltapi$ (s) &\multicolumn{2}{c}{ 75} &\multicolumn{2}{c}{300}   \\
 \hline
 $\ddpires$ (s)      &\multicolumn{2}{c}{ 0.5}&\multicolumn{2}{c}{3.1}  \\
 $\ddpiordre$ (s)      &\multicolumn{2}{c}{ 0.5}&\multicolumn{2}{c}{3.1}  \\
 $\ddpialias$ (s)      &\multicolumn{2}{c}{ 5.1}&\multicolumn{2}{c}{ 27}  \\
 \hline
 $\mathrm{max}(\specp)$&  15    & 30     & 15   & 30   \\
 $\ddpiover$ (s)       &  0.05  & 0.02   & 0.34 & 0.17 \\
\hline
\end{tabular}
\end{center}

\end{table}

\subsubsection{Uncertainties\label{Reliabilitylevel}}

The precision that can be achieved for the measurement of
$\deltapi$ depends on the time resolution of the Fourier spectrum
$\specp$ of the stretched spectrum. This resolution, related to
the properties of the oscillating signal, expresses as

\begin{equation}
   \ddpires = \numax\ \deltapi^{2}
   ,
   \label{equation-re}
\end{equation}
as derived in Appendix \ref{ap:resolution}. It then provides a
quantitative basis for estimating reliable uncertainties.

We investigated three different cases, depending on the mixed-mode
pattern density. If the mixed-mode pattern is dense with many
gravity-dominated mixed modes, then the precision on the
measurement is high since the function $\specp$  can be
oversampled (Fig. \ref{fig:TF}, top panel). It then writes, as a
function of the nominal resolution $\ddpires$,

\begin{equation}\label{eqt-erreur-S}
    \ddpiover \simeq {1.6\over A}\  \ddpires
    ,
\end{equation}
where $A$ is the maximum value of $\specp$ reached at $\deltapi$.

However, the accuracy on $\deltapi$ also depends on the constant
$\epsilong$ of the asymptotic gravity modes. At this stage, there
is no complete study on this parameter \citep[see][for a dedicated
study]{1986A&A...165..218P}, so that one cannot fix its value. An
uncertainty of 1 in $\epsilong$ translates into an uncertainty of
one radial gravity order. As shown in Appendix \ref{ap:ordre}, the
uncertainty is then

\begin{equation}
   \ddpiordre = \ddpires
   .
   \label{erreur_ordre2}
\end{equation}
This means that the high statistical precision has to be tempered
by our inability to determine the value of the offset $\epsilong$.
When only a low number of gravity dominated mixed-modes are
observed, it is not possible to unambiguously measure $\deltapi$
(second panel of Fig. \ref{fig:TF}). In this case, the uncertainty
corresponds to a shift of one radial order.

For evolved RGB stars, gravity-dominated mixed modes have
inertia that is too high, which means that they cannot be observed
\citep{2014A&A...572A..11G}. In such cases, the ambiguity for
measuring $\deltapi$ corresponds to a window effect. The absence
of observable mixed modes in the frequency ranges close to
quadrupole and radial modes yields large uncertainties. Instead, the observation of a few mixed modes in this region is
most often enough to remove any degeneracy in the solution. As
explained in Appendix \ref{ap:alias}, the frequency shift due to
missing gravity-dominated mixed modes around radial and quadrupole
modes is

\begin{equation}
   \ddpialias
   \simeq
   \nmax \ \ddpiordre
   ,
   \label{erreur_hole}
\end{equation}
and helps estimate the large uncertainty introduced by an alias
mismatch. For red giants, values of $\nmax$ are typically about
ten. Typical values of the uncertainties are given in
Table~\ref{table:incertitude}.

\section{Comparison with previous results\label{comparaison}}

To test the efficiency of the method, we compared the results obtained to
the \nstarM\ stars where \citet{2014A&A...572L...5M} have manually
measured the parameter $\deltapi$. We excluded subgiants and early red giants that have a large
separation $\Dnu$ larger than 18\,$\mu$Hz from their original data set. The oscillation
spectrum of these stars can be retrieved in short-cadence time
series only and are out of reach of the long-cadence data used
here.

\begin{figure}                 
  \includegraphics[width=9cm]{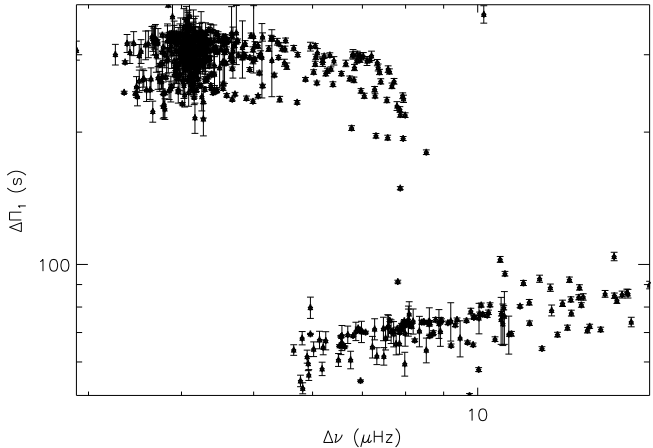}
  \caption{$\deltapi$ (s) in function of the large separation $\Dnu$ ($\mu$Hz)
  for the set of stars analyzed by \citet{2014A&A...572L...5M}. The uncertainties
  correspond to the 1-$\sigma$ error bars.}
  \label{fig:diagramme}
\end{figure}
\begin{figure}                 
  \includegraphics[width=9cm]{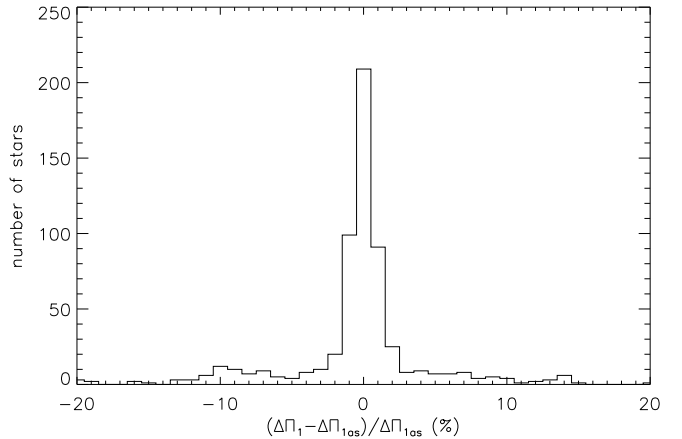}
  \caption{Histogram of the relative differences between the values $\deltapi$ obtained automatically or
  individually.}
  \label{fig:verif}
\end{figure}

We were able to deduce the $\deltapi$ for more than \accord\
stars (Fig. \ref{fig:diagramme}). The results show a good
agreement between the $\deltapi$ measured either manually or
automatically as shown in Fig.~\ref{fig:verif}, since the relative
difference is less than 2\,\% for more than 80\,\% of the stars,
and less than 10\,\% for more than 90\,\% of the stars. In fact,
the bump present at $-$10\,\% and $+$10\,\% corresponds to the
window effect described in Section~\ref{Uncertainties}. We
confirmed that a confusion between the different evolutionary
states is very rare. We found only two cases in this sample where
this situation was observed.

We note, however that the automated method fails to retrieve a
$\deltapi$ estimate in three main cases: \\
- Measuring $\deltapi$ is difficult at low $\Dnu$ when the number
of gravity-dominated mixed modes is small
\citep{2009A&A...506...57D,2014A&A...572A..11G} and when the
frequency resolution is poorer than the frequency difference
between consecutive mixed modes. This is particularly true for
evolved RGB
stars or AGB stars. \\
- When mixed modes have a low
visibility \citep{2012A&A...537A..30M,2014A&A...563A..84G}, only
the manual inspection of such stars can provide $\deltapi$, under
the condition that the mode visibility is not too small.\\
- In a limited number of cases, buoyancy glitches that induce a
modulation of the period spacing are large enough to hamper the
measurement of the mean period spacing with a Fourier analysis.
Often, the modulation is small and the method works, but
may deliver preferably an alias of $\deltapi$. We leave the analysis
of glitches to a forthcoming work.

On the contrary, we noted that the presence of rotational
splitting does not affect the determination of the $\deltapi$
value because the $\zeta$ function is
efficient at straightening the mixed-mode pattern even when
split by rotation. Each azimuthal order $m$ forms a family with
evenly spaced stretch periods, with a spacing ${\deltapi}_{,m}$ given by
\citet{2015A&A...584A..50M}

\begin{equation}\label{eqt-spacing-rot2}
    {\deltapi}_{,m}
    \simeq
    \deltapi \ \left( 1 + 2 m {\nmix\over \nmix+1} \, {\dnurot \over \numax} \right)
    ,
\end{equation}
where $\nmix$ (equal to $\Dnu / \deltapi \numax^2$) represents the
number of gravity modes in the $\Dnu$-wide frequency range around
$\numax$, and where $\dnurot$ is the maximum rotational splitting.
These rotational splittings are small compared to the frequency
$\numax$, so that the correction proportional to the azimuthal
order $m$ is in fact smaller than the resolution $\ddpires$
(Eq.~\ref{equation-re}). As a consequence, the period spacings of
all components of the dipole modes are close to $\deltapi$, and
rotation is not an issue for measuring $\deltapi$.
\newline

\begin{figure*}                 
  \includegraphics[width=9cm]{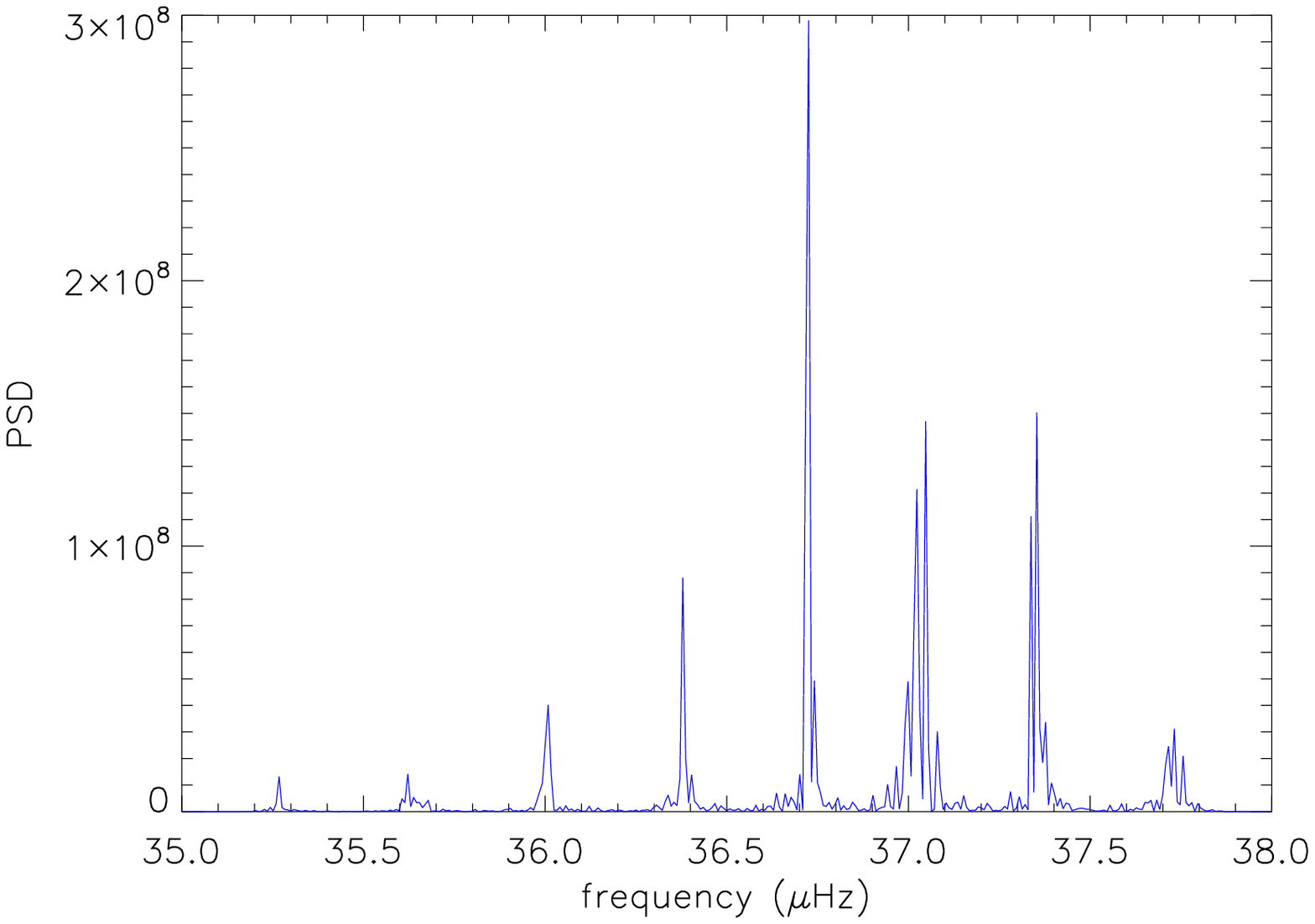}
  \includegraphics[width=9cm]{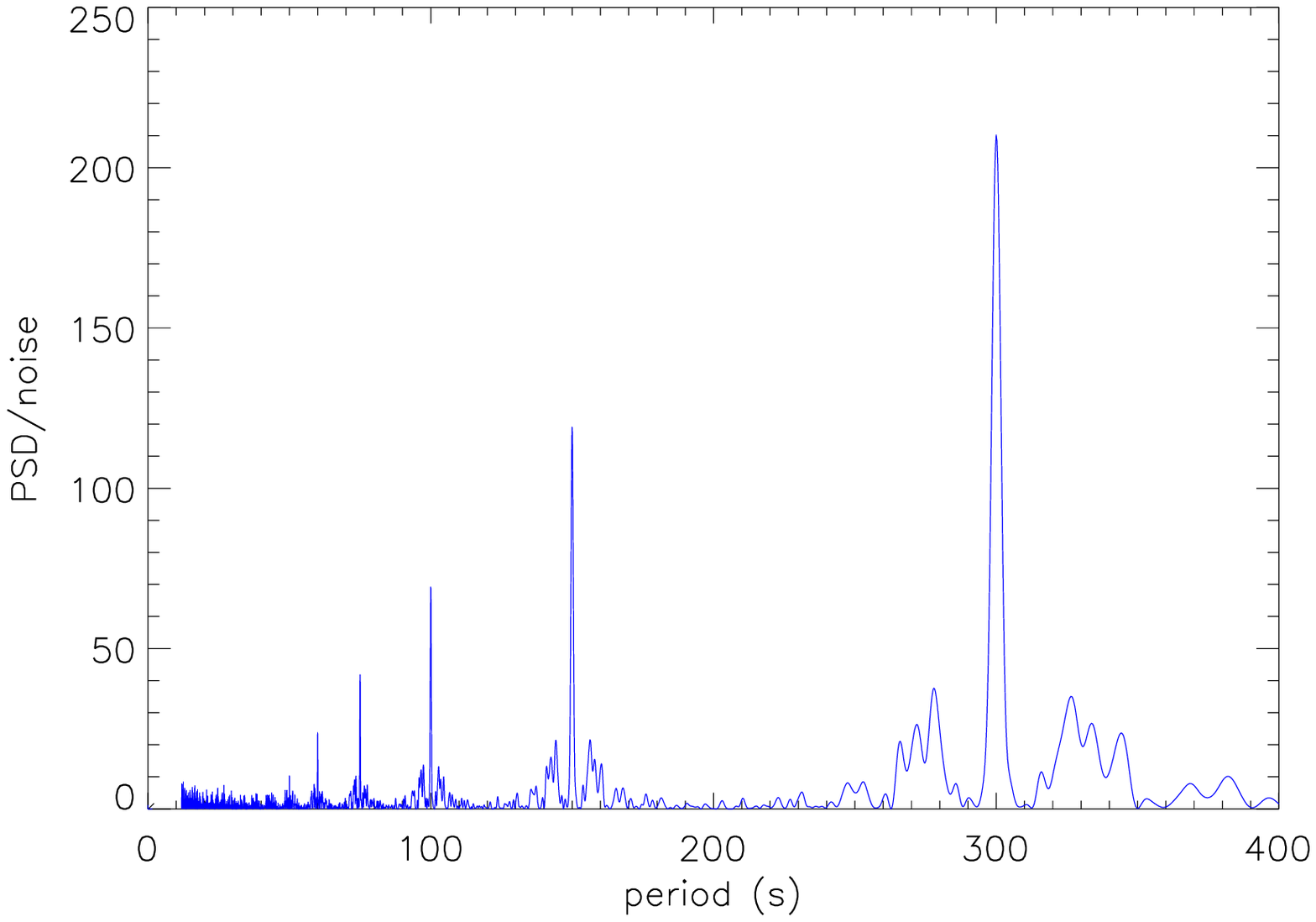}
  \includegraphics[width=9cm]{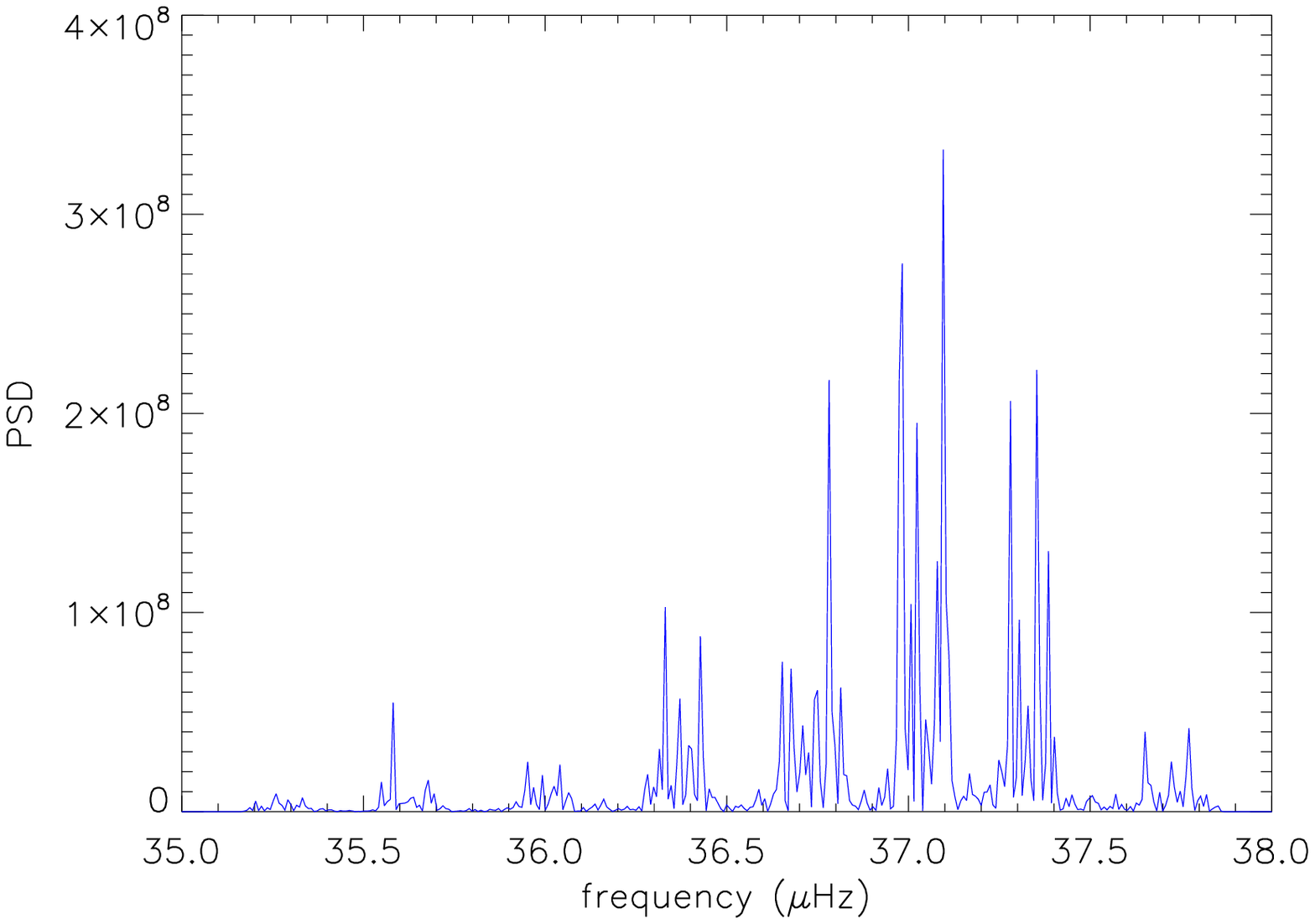}
  \includegraphics[width=9cm]{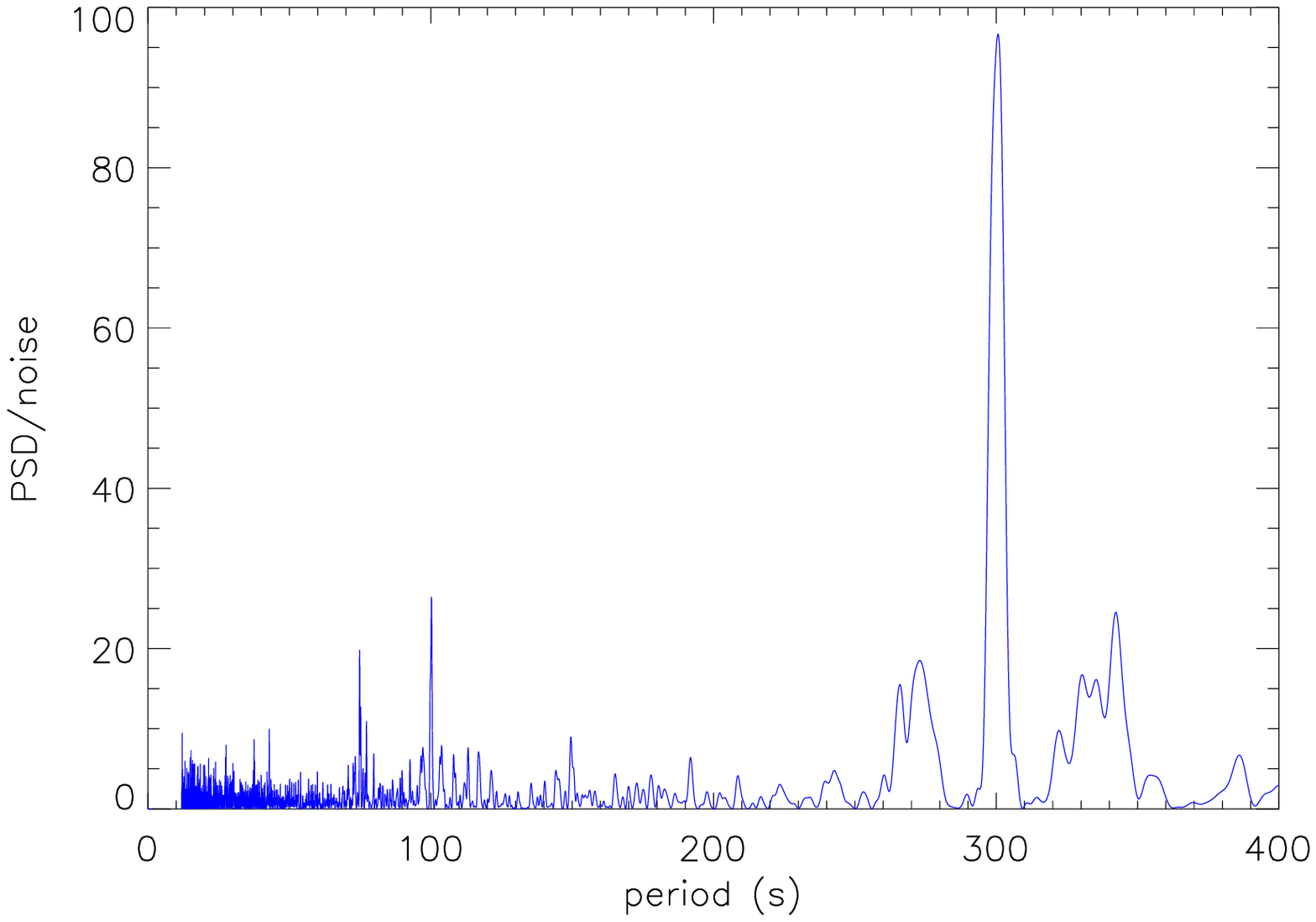}
  \caption{\textit{Left:} zoom on a precise radial order of the simulated mixed-mode spectrum shown in Figure \ref{simu}. $\chi^{2}$ noise with two degrees of freedom has been included in the spectrum with an height corresponding to $1/50$ of the height of the oscillations. \textit{Right:} Power spectrum of $\PPerdeb$ derived from the simulated spectra. \textit{Top:} simulations without rotation. \textit{Bottom:} same simulations but with the addition of rotational splittings, which appear as rotational triplets with a $\dnurot$ equal to $0.054$ $\mu$Hz.}
  \label{fig:spectre_rot}
\end{figure*}


In order to highlight these statements, we constructed several synthetic spectrum as described in Section \ref{Test_synthetic} with the addition of rotational splittings obtained from Eq. (17) of \citet{2015A&A...580A..96D}. We considered the case of a star seen with an inclination angle of 45$^\circ$ and a star seen equator-on corresponding respectively to the observation of rotational triplets and rotational doublets. In each case, the splitting amplitudes are considered equal. An example of a part of the synthetic spectrum is shown on the left side of Fig. \ref{fig:spectre_rot}. The power spectra of $\PPerdeb$ derived from the synthetic spectra, where rotational triplets were included, are shown on the right side of Fig. \ref{fig:spectre_rot}. Different values of $\deltapi$ and $\dnurot$ have been tested. In each case, the initial $\deltapi$ value has been retrieved with a precision higher than $0.2$ $\%$. The only visible signature of rotation is an apparent decrease in the power spectrum of $\PPerdeb$.

\section{Treatment of \Kepler\ red giant public data\label{traitement}}

\subsection{Data}

\begin{figure*}                 
  \includegraphics[width=18cm]{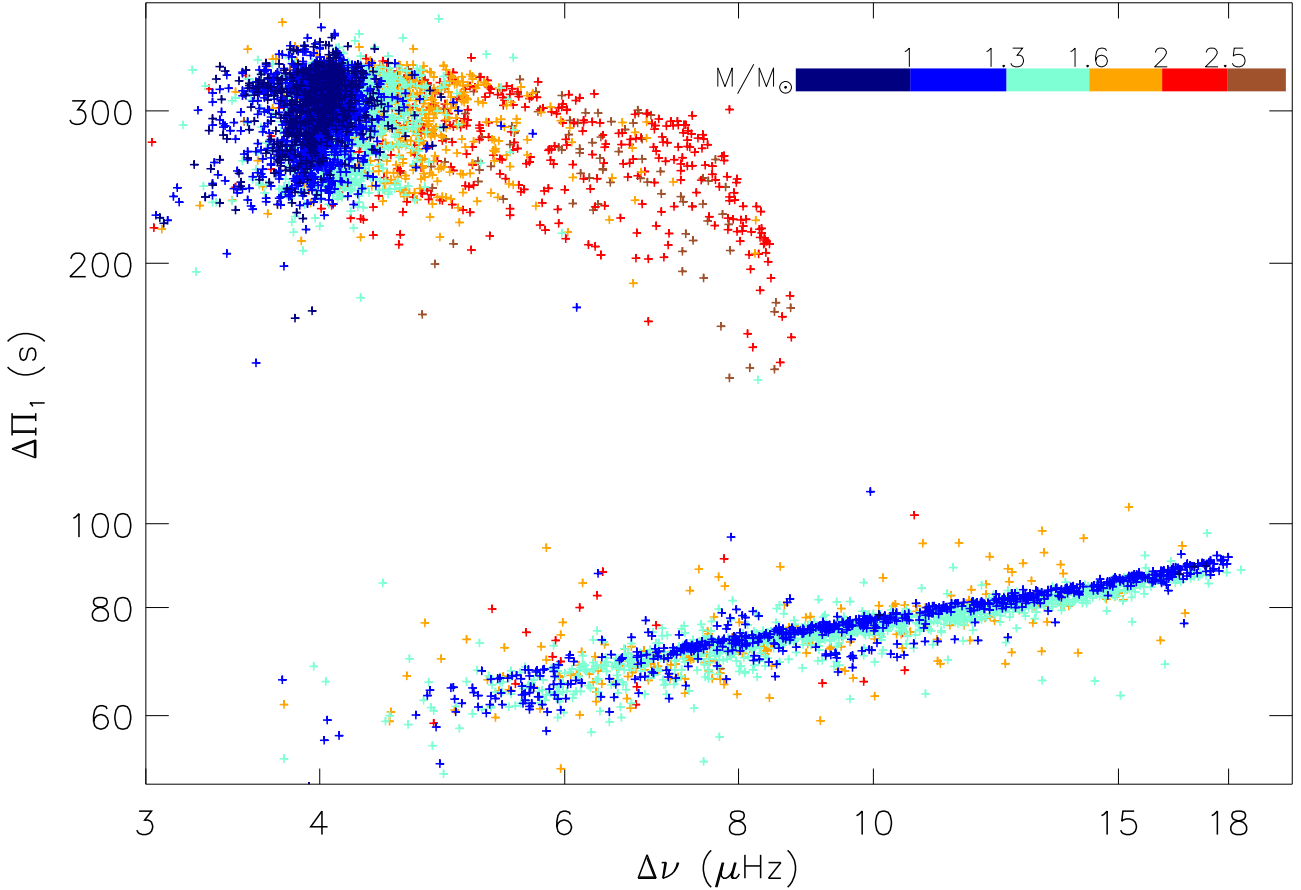}
  \caption{$\deltapi$ in function of the large separation $\Dnu$ for the
  \Kepler\ red giant public data. The color code indicates the stellar mass ($\Msol$)
  The signatures of the RGB, the main clump and
  the secondary clump  appear clearly. The proportion of outliers, compared to the seismic
  evolutionary tracks defined in \cite{2014A&A...572L...5M}, is less
  than 1\,\%.}
  \label{fig:deltapi-dnu}
\end{figure*}

We used the public long-cadence data from \Kepler\ with the
maximum available length, up to the quarter Q17, corresponding to
44 months of photometric observation. Original light curves were
taken from the MAST program \citep{2011KeplerF,2014KeplerF}. Among the 15\,000 light curves in the
\Kepler\ public data, we processed the set of more than 12900
stars for which the large separation $\Dnu$ can be reliably
measured.

\subsection{Gravity period spacing}

We were able to determine $\deltapi$ for about \detections\ red giants.
The other stars did not satisfy the reliability level defined in
Section \ref{Reliability level} owing to a low
signal-to-noise ratio in the spectra or to the presence of too few
g-dominated mixed modes. In some limited cases, it was due to
an incorrect identification of the radial mode pattern. Despite
the various checks, there are about 45 outliers in Figure
\ref{fig:deltapi-dnu}; they represent less than 1\,\% of the total
of the detections. We note, however, that large uncertainties due
to the alias problem affect about 20\,\% of the values, especially
on the RGB at low $\Dnu$. The data, obtained with two different codes, are available at the CDS (Table \ref{tab:CDS_file})

\begin{table*}[t]
\caption{Global seismic parameters \label{tab:CDS_file}}

\begin{center}
\begin{tabular}{cccccccccc}
 \hline
KIC number & $\Dnu$ ($\mu$Hz) & $\deltapi$ (s) & $\delta\deltapi$ (s) & $q$ & $M/M_\odot$   & $\delta M/M_\odot$ & alias & method & evolution\\
\hline
1868101  &  3.80   &  298.7   &   2.80   &   0.35   &   0.93   &   0.05   &    0    &   2   &   1\\
1995859  &  4.74   &  321.4   &   4.71   &   0.29   &   1.86   &   0.06   &    0    &   2   &   1\\
9145955  & 11.03   &   77.1   &   1.50   &   0.16   &   1.44   &   0.06   &    0    &   1   &   0\\
12507577 &  7.52   &   69.7   &   0.40   &   0.21   &   1.45   &   0.06   &    0    &   2   &   0\\
\hline
\end{tabular}

\end{center}

Global seismic parameters, as described in the text, of the four stars
shown in Figs.~\ref{fig:spectre_debump} and \ref{fig:TF}. The full table
corresponding to the whole data set shown in Fig. \ref{fig:deltapi-dnu}, with about \detections\ red giants, is
available at CDS via anonymous ftp to cdsarc.u-strasbg.fr (130.79.128.5) or via http://cdsarc.u-strasbg.fr/viz-bin/qcat?J/A+A/NNN/PPP". 
Uncertainties on $\deltapi$ are estimated according
to the expressions given in the Appendix. They depend on the robustness
of the measurement, which is indicated in the columns 'alias'and 'method'
The 'alias' value is set to 1 if the value of $\deltapi$ likely
corresponds to that of an alias, otherwise to 0.
The 'method' value indicates whether 1 or 2 codes using the method
presented in the article could converge and provide a relevant value for
$\deltapi$; a value of 2 also means that both codes provide consistent
results.
The 'evolution' value provides the evolutionary stage: 0 for RGB stars,
1 for clump stars, or 2 for secondary clump stars.

\end{table*}

The results confirm and extend the conclusions of
\citep{2011Natur.471..608B,2012A&A...540A.143M,2013ApJ...765L..41S,2014A&A...572L...5M}, taking into account that here we measure
asymptotic values and not mean period spacings. We see the same
characteristic features except for stars starting the ascension of
the AGB identified by \citet{2014A&A...572L...5M}. Their absence
can be explained by the low signal-to-noise ratio of oscillation
spectra at low $\Dnu$, which then induces the rejection of the
measurements; instead, such stars can be analyzed
individually.

\begin{table}[t]
\caption{Values of the period spacings $\deltapi (M,Z)$, in seconds, for a large separation $\Dnu = 6\,\mu$Hz, with the scaling relation $\deltapi \propto \Dnu^{0.25}$. Uncertainties are of about 0.25\,s. \label{tab:fitMZ}}

\begin{center}
\begin{tabular}{ccccc}
 \hline
$M/M_\odot$   & 1.0-1.2 & 1.2-1.4 & 1.4-1.6 \\
\hline
$Z < -0.4$    & 68.0  &  67.9  &  67.3 \\
$-0.4 < Z < 0$& 68.7  &  68.2  &  67.4\\
$ Z > 0$      & 69.1  &  68.5  &  67.5\\
\hline
\end{tabular}
\end{center}

\end{table}

\subsection{Mass and metallicity}

\begin{figure*}
  \includegraphics[width=18cm]{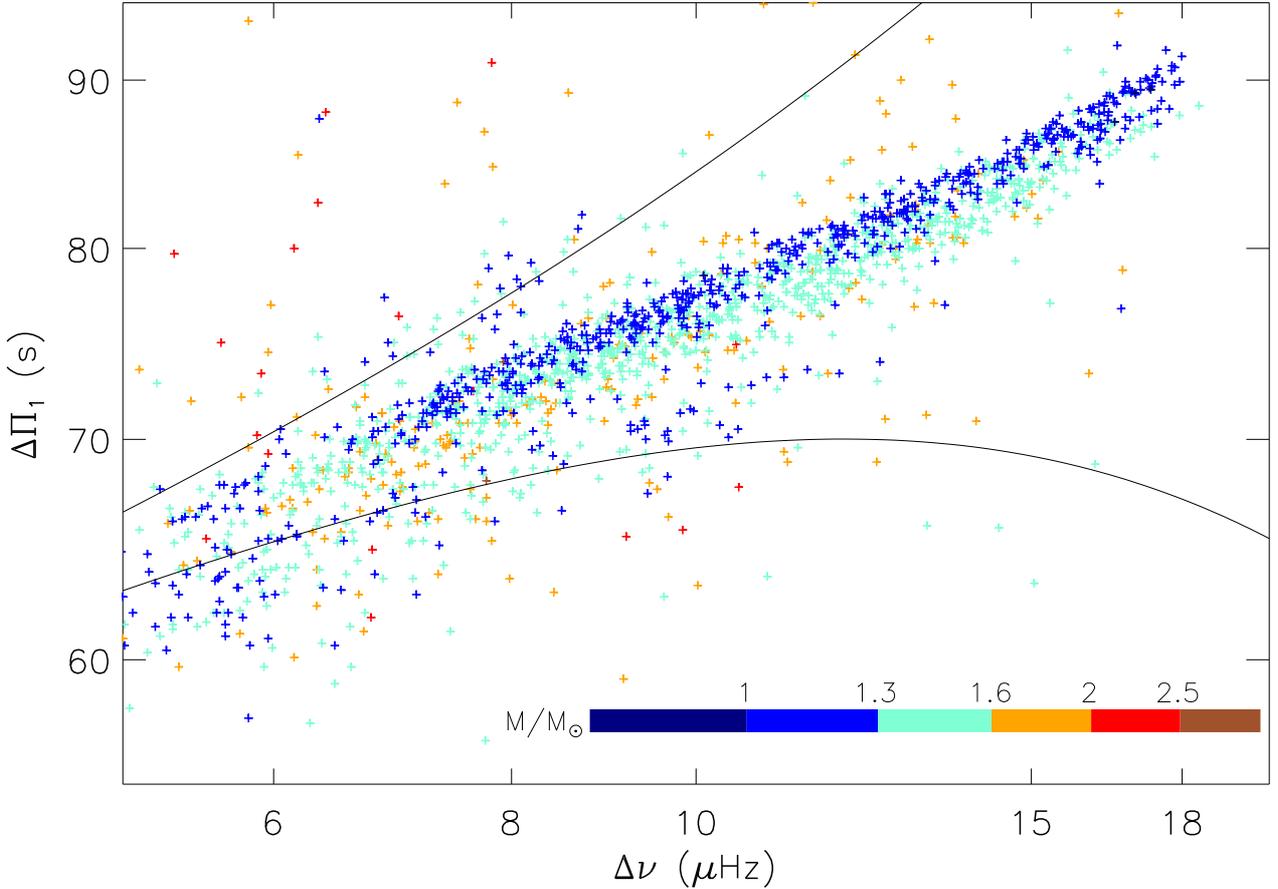}
  \caption{$\deltapi$ (s) in function of the large separation $\Dnu$ ($\mu$Hz)
  for the \Kepler\ red giant public data. The color code indicates the stellar mass ($\Msol$).
 We have superimposed two curves indicating the regions where aliases are
  expected. We note the large number of such artefacts below 9\,$\mu$Hz.}
  \label{fig:RGB_masse}
\end{figure*}

\begin{figure}
  \includegraphics[width=9cm]{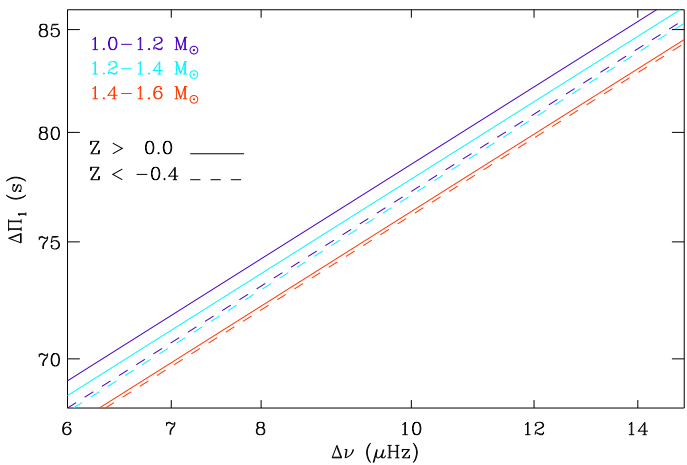}
  \caption{Fit of the $\deltapi$ -- $\Dnu$ relation depending on the stellar mass and
  metallicity.
  }
  \label{fig:RGB_masse-metal}
\end{figure}

\begin{figure*}
  \includegraphics[width=18cm]{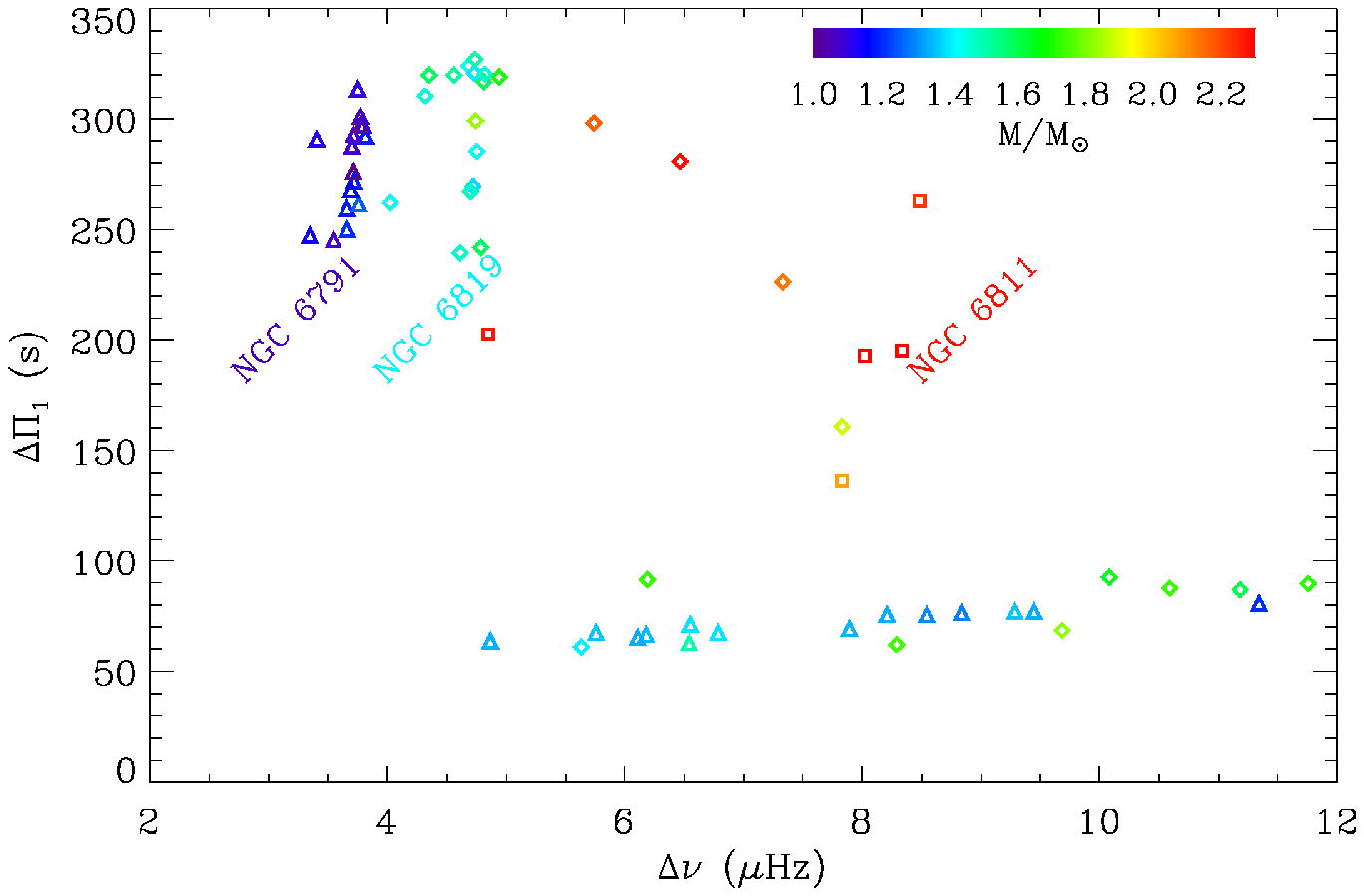}
  \caption{Same as Fig. \ref{fig:deltapi-dnu}, with red giants
  identified as members of three open clusters observed by
  \Kepler : NGC 6791, triangles; NGC 6819, diamonds; NGC 6811,
  squares.
  }
  \label{fig:clusters}
\end{figure*}

The scaling relations allow the determination of the stellar masses and radii from the global seismic parameters ($\Dnu$, $\numax$) and from the effective temperature \citep[e.g.,][]{2010A&A...522A...1K}. We used the effective temperature listed in \citet{2014ApJS..211....2H}. The relative uncertainties on the stellar mass are about 10-15$\%$. The variations of $\deltapi$ along the stellar evolution for different star masses can then be analyzed.

The mass dependence present in the main and secondary clumps \citep[e.g.,][]{2011Natur.471..608B,2014A&A...572L...5M} is confirmed.
This is illustrated in Fig.~\ref{fig:clusters} with a
subsample of Fig.~\ref{fig:deltapi-dnu}, with stars from the open
clusters
 NGC 6791 ($M\ind{6791} = 1.15 \pm 0.03 \,M_\odot$ in the RGB),
 NGC 6811 ($M\ind{6811} = 2.2 \pm 0.1\,M_\odot$),
 and
 NGC 6819 ($M\ind{6811} = 1.61 \pm 0.04 \,M_\odot$).
The membership of these stars is defined as in
\cite{2010ApJ...713L.182S}; the masses are derived from
\cite{2012MNRAS.419.2077M}. The evolutionary tracks of the three
clusters are close to each other on the RGB, where the mass
dependence is weak, but important in the helium-burning phase. We
confirm the identification of three blue stragglers in NGC 6819,
already identified by \cite{2012ApJ...757..190C}.

We also discovered another mass dependence present in the RGB
branch. At fixed properties of the stellar core (at fixed
$\deltapi$), a high-mass RGB star has a larger $\Dnu$ value than a
low-mass star. This means that, despite a more massive convective
envelope, high-mass stars are more dense. This relation, observed
for RGB stars with a mass below 1.6\,$\Msol$, is predicted by
simulations \citep{2013ApJ...765L..41S}, but has never been observed. We also
note that stars on the RGB with a mass above 1.6\,$\Msol$
exhibit a large spread around the RGB branch. This phenomenon,
already noted by \citet{2014A&A...572L...5M}, can be related to
the different physical conditions when such stars reach the RGB.

The large number of stars on the RGB with a precise measurement of
$\deltapi$ also allowed us to test the metallicity dependence of
the $\deltapi$ -- $\Dnu$ relation. At fixed
properties of the core (at fixed $\deltapi$), we observe that low metallicity
stars have a large spacing $\Dnu$ significantly higher than most
metallic stars (Fig.~\ref{fig:RGB_masse-metal}). The values of
$\deltapi (M,Z)$ for $\Dnu=6\,\mu$Hz are given in
Table~\ref{tab:fitMZ}. Uncertainties on these values are about
0.25\,s, significantly less than the observed spread. The metallic
dependance is high for stars below 1.2\,$M_\odot$, but negligible
for stars above 1.4\,$M_\odot$. This agrees with the fact that
low-metallicity stars are denser. The fits in
Fig.~\ref{fig:RGB_masse-metal} were obtained assuming that the
slope of the $\log(\deltapi)$ -- $\log(\Dnu)$ relation is fixed,
equal to 0.25 according to the global fit.

\subsection{Luminosity bump}

We also note the large spread on the $\deltapi$ value for RGB
stars with $\Dnu$ lower than $6.5$ $\mu$Hz. This spread occurs
where simulations predict the position of the luminosity bump
\citep{2012A&A...543A.108L}. The spread could then be due to such a
phenomenon.

The aliasing phenomenon complicates the result at low $\Dnu$ on
the RGB and  precludes the firm identification of the signature of
the luminosity bump, as discussed above. The main aliases are present
as the second branch observed under the RGB branch.

\begin{figure}                 
  \includegraphics[width=9cm]{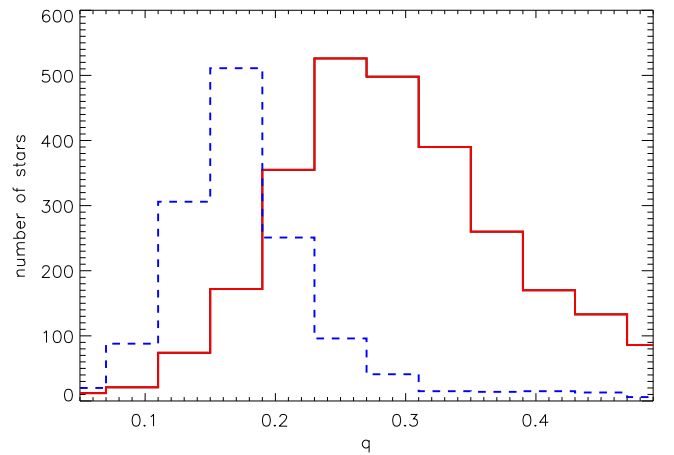}
  \caption{Histogram of the coupling factor $q$. The dashed blue line corresponds
  to RGB stars and the continuous red line to clump stars.
  \label{fig:q_result}}
\end{figure}

\subsection{Coupling parameter}

The coupling factor $q$ was adjusted along with $\deltapi$. The
results are shown in Fig. \ref{fig:q_result}.

Red giant branch stars show a smaller coupling than clump stars, as stated by
\citet{2012A&A...540A.143M}. Our results are similar to this study
for RGB stars: the mean value is $q =0.17 \pm 0.05$. However, the
results for clump stars are higher: around $0.29 \pm 0.07$, but
still in the previously estimated uncertainties. The clump stars present a
stronger coupling between p and g modes than RGB stars.

Following \cite{1989nos..book.....U}, the value of the coupling
factor is linked to the extent of the evanescent region and is
limited to 1/4. Our results, however, do not verify this: we measure
$q$ values significantly above 1/4. This discrepancy arises because the formalism in \cite{1989nos..book.....U} is only valid
for a weak coupling (M. Takata, private communication), which
is not observed. A new development is necessary to link $q$ and
the size of the evanescent zone, as stated by
\cite{2014MNRAS.444.3622J}.

\section{Conclusion\label{conclusion}}

We presented a new method based on the inversion of the mixed-mode
asymptotic relation for determining automatically the gravity
period spacing in the mixed-mode pattern of red giants stars. The
efficiency of the method derives from the asymptotic properties of
the radial and dipole spectrum: the radial oscillation pattern
follows the universal red giant oscillation pattern, which
corresponds to a second-order asymptotic development; the dipole
oscillation pattern is tightly fitted by the asymptotic expansion
for mixed modes. The change of variable used to analyze the
stretched periods of the modes allows us to exhibit the properties
of the oscillation periods and not of period spacings.

We used this new method on the red giant \Kepler\ public data and
succeeded in deducing the gravity period spacing for about
\detections\ red giants. The results obtained confirmed previous
measurement for these stars. We unveil as well a new mass
dependence for RGB stars: higher-mass stars will have a lower
period spacing. This work paves the way to the precise
interpretation of the mixed-mode pattern on numerous stars; a
massive measurement of rotational splittings is now possible. We
note that buoyancy glitches sometimes hamper the detection of
$\deltapi$, but such glitches deserve a specific
study.

\begin{acknowledgements}
We acknowledge the entire Kepler team whose efforts made these
results possible. We acknowledge financial support from the
Programme National de Physique Stellaire (CNRS/INSU) and from the
ANR program IDEE Interaction Des \'Etoiles et des Exoplan\`etes.
MV thanks K\'evin Belkacem, Thomas Kallinger, and Tim White for
fruitful discussions.
\end{acknowledgements}
\bibliographystyle{aa} 
\bibliography{dPi1_6.bib}

\appendix

\section{Uncertainty on $\deltapi$\label{ap:1}}

The measurement of $\deltapi$ depends on the characteristics of
the stretched spectrum and of its spectrum. We first
examine how they are connected with the method and with the global seismic
properties of the spectrum. The exact measurement of $\deltapi$
presupposes as well the exact determination of the mixed-mode
order, which depends on the value of the gravity offset
$\epsilong$. The measurement is also perturbed by the low
amplitudes of gravity-dominated mixed modes. Since these modes are
evenly spaced in frequency, this effect is comparable to a window
effect. All these effects are examined.

\subsection{Uncertainties of the stretching process\label{ap:erreur}}

The accuracy of the stretching process is ensured by the principle
of the method and the use of Eq.~(\ref{stretched}). In this
equation, the difference between periods and stretched periods is
due to the term $1/\zeta$, which basically ensure that there are
$\nmix +1$ modes in a $\Dnu$-wide interval, where $\nmix=\Dnu /
\deltapi\nu^2$ modes are expected. As a result, the relative
difference between periods and stretched periods is measured by
$1/\nmix$. Except for subgiants and on the early RGB where $\nmix$
has small values \citep{2014A&A...572L...5M}, the large values of
$\nmix$ for the RGB and clump stars considered in this work
ensures an efficient iteration. In order to illustrate this, Table
\ref{tab:iteration} shows the convergence of the iteration process
in the case of an RGB star with $\deltapi= 75$\,s, with two
initial guess values (RGB or clump). Even if the initial guess
value corresponds to an incorrect determination of the
evolutionary stage, the iteration process precisely converges
after four steps.

\subsection{Resolution of the spectrum of the stretched spectrum\label{ap:resolution}}

We examine the properties of the stretched spectrum and
of its spectrum.

Since the frequency range where modes are observed is typically
$\numax$, the period range of the stretched spectrum is about
$1/\numax$. Then, the spectrum of this spectrum typically has a
resolution $\numax$. Since the signature of the period spacing
$\deltapi$ occurs at $1/ \deltapi$, the nominal resolution
expressed in the period variable is

\begin{equation}
   \ddpires = \numax \deltapi^{2}.
   \label{erreur_resol}
\end{equation}
Typical values of the nominal resolution correspond to about
0.4\,s on the RGB and 3.6\,s in the red clump.

\subsection{Uncertainties with an oversampled spectrum}

The high quality of signal allows us to oversample the spectrum in
order to increase the resolution. This tighter resolution is,
however, limited by the noise. Following the same approach as in
\cite{2009A&A...508..877M}, and especially their equations
(A.4)-(A.6), we can compare the variation of the signal peaking at
amplitude $A$ to the maximum variation of a noise contribution of
amplitude $b$ (both expressed in white noise units). The precise
identification of the signal maximum allows us to compare the
oversampled resolution $\ddpiover$ to the nominal resolution

\begin{equation}\label{eqt-compar}
    \pi A \ \ddpiover \ge b \ \ddpires
    .
\end{equation}
Considering a conservative value $b=5$, in white noise units, we
have

\begin{equation}\label{eqt-compar2}
    \ddpiover \simeq {1.6\over A}\  \ddpires
    .
\end{equation}
Values of $A$ above the threshold level ensure a significantly
tighter resolution than the nominal value $\ddpires$. Since $A$
may be as high as 200, the accuracy of the measurement may become
excellent; however, this supposes that the gravity offset
$\epsilong$ intervening in the pure gravity-mode pattern is known.
It is usually set to 0, although the asymptotic value
is 1/4 \citep{1980ApJS...43..469T}. Anyway, its expression can be
complicate, depending on the structure of the radiative core
\citep{1986A&A...165..218P}.

\subsection{Uncertainties corresponding to a shift of one gravity order\label{ap:ordre}}

According to the asymptotic theory, the gravity mode periods are
evenly spaced with a mean period spacing close to $\deltapi$. So,
their frequencies express

\begin{equation}
   \nu = \frac{1}{\ng\deltapi}
   ,
   \label{deltapi_regulier}
\end{equation}
where $\ng$ is the gravity order. By deriving this equation, we
obtain

\begin{equation}
   \frac{\diff\nu}{\nu} = - \frac{\diff\ng}{\ng} - \frac{\diff\deltapi}{\deltapi}  .
   \label{deltapi_regulier_derivee}
\end{equation}
A shift of one radial order corresponds to $\diff \ng = 1$. For a
fixed set of  oscillation modes, we have $\diff\nu=0$. It follows that
$\diff\deltapi = \deltapi / \ng$. By assuming that we are near the
maximum oscillation frequency $\numax$ and following Eq.
(\ref{deltapi_regulier}), the error on $\deltapi$ can be written
as

\begin{equation}
   \ddpiordre = \numax \deltapi^{2}
   .
   \label{erreur_ordre}
\end{equation}
So we note that we obtain a similar result to that determined by the
resolution:

\begin{equation}
   \ddpiordre = \ddpires
   .
   \label{erreur_erreur}
\end{equation}

\subsection{Uncertainties corresponding to the window effect\label{ap:alias}}

The suppression of the radial and quadrupole modes we performed
possibly leads to what is called a window effect. The removal of
part of the spectrum at regular frequencies produces aliases
(equal to $1/\Dnu$ in period), which could be mistakenly attributed
to the real value of $\deltapi$. To estimate the uncertainties
related to this possible confusion, we also have to estimate the
frequency difference between each mode. To this end, we follow Eq.
(\ref{deltapi_regulier_derivee}) with a fixed $\deltapi$ value
($\diff\deltapi = 0$) and an order variation $\diff\ng=1$:

\begin{equation}
   \frac{\diff\nu}{\nu} = - \frac{1}{\ng}
   .
   \label{intermede}
\end{equation}
With $\ng$ determined by  Eq.~(\ref{deltapi_regulier}), the value
of $\diff\nu$ is equal to $\deltapi\nu^{2}$. It corresponds to a
signature at the period $1/(\deltapi\nu^{2})$. The relative period
shifts due to the aliases are then determined by

\begin{equation}
   \frac{\diff\deltapi}{\deltapi}
   =
   { \displaystyle{1\over \Dnu}
     \over
     \displaystyle{1 \over {\deltapi\nu^{2}}}
   }
   .
   \label{deltapi_erreur_rapport}
\end{equation}
For frequencies close to the maximum oscillation frequency
$\numax$, this shift translates into an uncertainty on $\deltapi$
expressed by

\begin{equation}
   \ddpialias = \frac{(\numax\deltapi)^{2}}{\Dnu}
   .
   \label{deltapi_erreur_trou}
\end{equation}
This, however, relies on the detection of pure gravity modes. Here,
mixed modes with stretched periods behave as gravity modes, but
with an additional mode that is the pressure mode. Hence, the
correct uncertainty is reduced by a factor $\nmix / (\nmix +1)$ so
that

\begin{equation}
   \ddpialias
   =
   {(\numax\deltapi)^{2}
   \over
   \Dnu}
   {\nmix \over \nmix+1}
   =
   {\deltapi
   \over \nmix+1}
   .
   \label{deltapi_erreur_trou2}
\end{equation}

\begin{table}[t]
\caption{Iteration process\label{tab:iteration}}
\begin{tabular}{ccc}
\hline
  Iteration step $i$&  \multicolumn{2}{c}{${\deltapi}_i$ (s)} \\
\hline
 start value  & 80     &   300    \\
       1      & 75.30  & 88.50    \\
       2      & 75.018 & 75.810   \\
       3      & 75.0011& 75.0486  \\
       4      & 75.0001& 75.0029  \\
 target value & 75     & 75       \\
\hline
\end{tabular}
\end{table}

\end{document}